\documentclass[12pt]{article}
\usepackage[margin=1in]{geometry}
\usepackage{booktabs,graphicx,amsmath,amssymb}
\usepackage{caption}
\usepackage{float}
\usepackage[hidelinks]{hyperref}
\usepackage[round,authoryear]{natbib}
\usepackage{setspace}
\usepackage{threeparttable}
\graphicspath{{./}}

\title{\textbf{Helping Hands, Healthier Infants:\\ The Effect of Medicaid Doula Coverage Mandates on Birth Outcomes}}
\author{Farhad V. Farahani\\ University of South Florida}
\date{Third Year Term Paper \\[4pt] \today}

\begin{document}
\maketitle

\begin{abstract}
\noindent Over the last decade a wave of U.S.\ states began reimbursing doula
services through Medicaid, hoping to improve infant health and narrow stark
racial gaps in birth outcomes. I evaluate these mandates using the staggered
2021--2024 rollout, a panel of 32.1 million births from CDC WONDER (2016--2024),
and a newly assembled measure of the state doula workforce drawn from the
national provider registry. Identification comes from the policy's timing rather
than from comparing doula users to non-users, addressing the selection problem
that limits the existing observational literature. On average I find no
detectable effect on low birth weight (LBW). Consistent with the heterogeneity
emphasized by \citet{peet2022variation} and the maternal-health-disparities
literature, however, the effect concentrates among the group at greatest risk:
Black mothers, for whom LBW falls by roughly half a percentage point
($\approx$5\% of the baseline) in the states with the longest exposure, with flat
pre-trends and a coherent upward shift in the birth-weight distribution. The
estimate is marginal once I use inference valid for few treated clusters, and the
binding constraint is statistical power: most mandates took effect in 2024--2025,
at or beyond the end of the data. A two-stage least squares analysis
shows that coverage roughly doubles the doula workforce (first-stage
$F\approx21$--$35$), and that the induced increase in doula supply is associated
with lower Black LBW, though imprecisely. I read the results as credible early
evidence that doula mandates work where they have had time to operate and where
the need is greatest, rather than as a finished causal claim.

\medskip
\noindent\textbf{Keywords:} doula care; Medicaid; low birth weight; racial
disparities in infant health; difference-in-differences; maternal health policy.

\smallskip
\noindent\textbf{JEL codes:} I18, I14, I13, J13, J15, C21.
\end{abstract}

\thispagestyle{empty}
\newpage
\setcounter{page}{1}
\onehalfspacing

\section{Introduction}
\begin{figure}[H]
\centering
\includegraphics[width=0.86\linewidth]{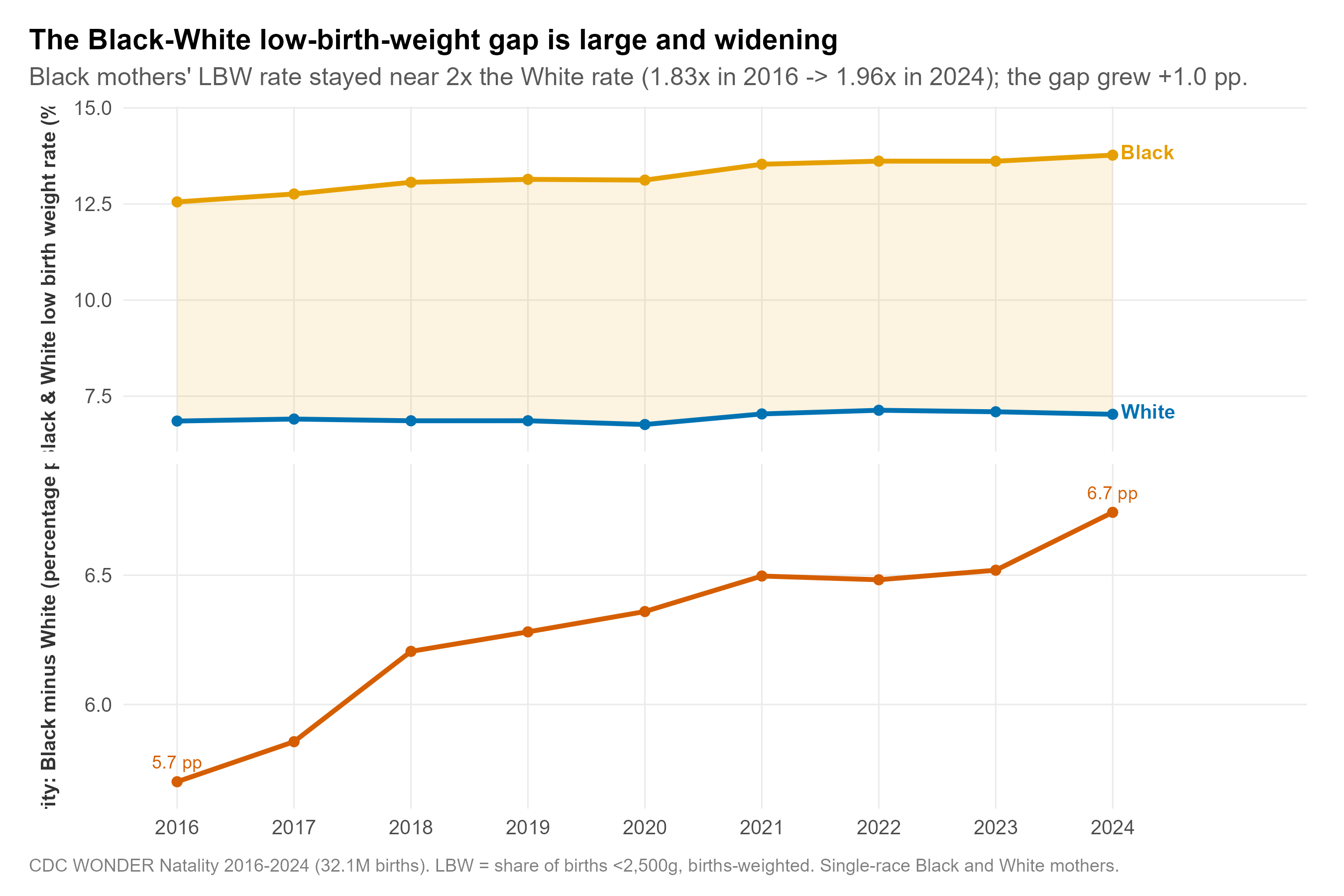}
\caption{The Black--white low-birth-weight gap is large and widening, 2016--2024.
Top: LBW rates by maternal race, with the gap shaded. Bottom: the gap in
percentage points. \emph{This is the disparity the policy targets and the
motivation for the paper's focus on Black mothers.}}
\label{fig:gap}
\end{figure}

Black infants in the United States are born too small far more often than white
infants, and the gap has not closed. Over 2016--2024 the low-birth-weight (LBW)
rate among births to Black mothers averaged 13.3\%---nearly one in seven---against
7.0\% for white mothers, a ratio that \emph{widened} from 1.83 to 1.96 over the
period (Figure~\ref{fig:gap}). Because birth weight is among the most reliable
early predictors of later health, schooling, and earnings
\citep{aizer2014intergenerational}, this gap is both a health and an economic
problem, and it has made maternal health disparities a first-order policy concern.

One increasingly popular response is the doula: a trained, non-clinical companion
who provides continuous emotional, informational, and physical support before,
during, and after birth. A large clinical literature, summarized by the Cochrane
review of \citet{bohren2017continuous}, finds that continuous labor support,
relative to usual care, improves birth experiences and reduces medical
intervention, with the largest benefits when the support person is neither
clinical staff nor a member of the woman's own social network---that is, a doula. Until recently, however, doula care was
largely an out-of-pocket service, effectively rationed away from the low-income
and disproportionately Black mothers who might benefit most. Between 2021 and
2025, more than two dozen states moved to cover doula services through Medicaid,
creating---for the first time---broad policy variation with which to study the
question at scale.

This paper asks whether these Medicaid \emph{doula coverage mandates} improved
birth outcomes, and for whom. The central empirical challenge is selection. The
existing evidence---influential observational studies such as
\citet{kozhimannil2013doula} and \citet{falconi2024role}, and the
machine-learning evaluation of \citet{peet2024infant}---compares mothers who
\emph{used} a doula to those who did not. Because doula use is chosen, these
comparisons risk conflating the effect of doulas with the characteristics of
women who seek them out. Whether doula coverage should be expanded, and to which
populations, depends on the \emph{causal} effect of the program itself, not on the
characteristics of the women who currently choose doula care; a correlation driven
by selection cannot answer that policy question. I take a different approach: I
exploit the staggered \emph{timing of state mandates} in a
difference-in-differences (DiD) design, so that identification rests on policy
variation rather than on individual take-up.

I combine three sources of data: the universe of U.S.\ births by state, year,
race, education, and birth-weight bin from CDC WONDER (2016--2024); hand-coded
effective dates of statewide Medicaid doula coverage; and a state-by-year measure
of the doula workforce built from 19{,}425 individual doula records in the
national provider registry. I estimate average effects with both two-way
fixed-effects (TWFE) and the heterogeneity-robust estimator of
\citet{callaway2021difference}; I conduct inference appropriate for a small number
of treated clusters using the wild cluster bootstrap
\citep{cameron2008bootstrap,mackinnon2017wild} and randomization inference; and I
use coverage as an instrument for doula supply.

The findings are easy to summarize.
First, the \emph{average} effect of coverage on LBW is a precise zero. Second,
and following directly the logic of \citet{peet2022variation}---that average
effects can ``mask important benefits to certain groups,'' and that programs are
``most beneficial among those at greatest risk''---the effect is concentrated
among Black mothers. In the earliest-adopting states, LBW among Black mothers
falls by about 0.52 percentage points (roughly 5\% of baseline); the pre-trends
for this group are flat, the effect grows with years of exposure, and it shows up
as a coherent shift of mass out of the low-birth-weight range and into the
normal range. Third, this result is best described as a \emph{marginal,
suggestive} one: with only five treated states carrying meaningful post-period
exposure, valid few-cluster inference puts the two-sided $p$-value near $0.10$
(one-sided $\approx 0.03$). Fourth, the mechanism is visible at its first link:
coverage roughly doubles the doula workforce.

The binding constraint throughout is statistical power. Fourteen of the
twenty-seven adopting states began coverage in 2024--2025, at or after the end of
the birth data, so most ``treated'' states contribute almost no post-treatment
observations. This is a feature of the policy timeline, not of the method, and it
means the present estimates should firm up---or fade---as 2025--2026 natality data
arrive.

The paper contributes to two literatures. To the doula and maternal-health
literature \citep{kozhimannil2013doula,kozhimannil2016modeling,falconi2024role,
peet2024infant,mottlsantiago2023effectiveness}, it adds the first
quasi-experimental (policy-variation) evaluation of statewide mandates,
complementing the largely descriptive, selection-prone evidence from comparisons
of doula users to non-users. I emphasize that this design, while it exploits
policy timing rather than individual choice, is still observational rather than a
randomized trial: its credibility rests on the parallel-trends assumption, not on
random assignment. To the health-economics literature on perinatal policy
\citep{guldi2023effects,sonchak2015medicaid,cyganrehm2022effects,
noghani2023longterm,anderson2020occupational}, it adds evidence on a new and
rapidly spreading instrument---doula coverage---and, like \citet{sonchak2015medicaid}
and \citet{anderson2020occupational}, finds the action concentrated among
disadvantaged and Black mothers.

The remainder of the paper proceeds as follows. Section~\ref{sec:lit} reviews the
related literature. Section~\ref{sec:data} describes the policy setting and data.
Section~\ref{sec:strategy} lays out the empirical strategy and the identification
threats. Section~\ref{sec:results} presents results: the average null, the
heterogeneity that is the heart of the paper, inference, and the mechanism.
Section~\ref{sec:discussion} discusses limitations, and Section~\ref{sec:conclusion}
concludes.

\section{Related Literature}\label{sec:lit}
\paragraph{Do doulas work?} The clinical foundation is the Cochrane review of
\citet{bohren2017continuous}, pooling 27 randomized trials of roughly 16{,}000
women: relative to usual care, continuous labor support increases spontaneous
vaginal birth (risk ratio $1.08$, 95\% CI $1.04$--$1.12$), lowers cesarean
delivery (RR $0.75$, $0.64$--$0.88$), and reduces negative ratings of the birth
experience (RR $0.69$, $0.59$--$0.79$), with the largest effects when the support
person is a doula rather than clinical staff or the woman's own network, and no
evidence of harm. The randomized trial of \citet{mottlsantiago2023effectiveness}
extends this to racially and culturally congruent community doulas in a U.S.\
safety-net hospital ($n=367$). Overall differences were small, but---strikingly
for the present paper---benefits concentrated among Black birthing people, who
experienced a (statistically insignificant) 12.9 percentage-point reduction in
cesarean birth and an 11.5 percentage-point increase in exclusive breastfeeding,
leading the authors to call for larger studies of community doula support for
Black mothers.

\paragraph{Doulas and Medicaid.} \citet{kozhimannil2013doula} provided the
foundational Medicaid evidence, finding markedly lower cesarean and preterm rates
among doula-supported Medicaid births; \citet{kozhimannil2016modeling} showed
such coverage could be cost-neutral or cost-saving. The most recent national
study, \citet{falconi2024role}, uses propensity-matched Medicaid claims (2014--2023)
to estimate 29\% lower preterm risk among doula users. These studies are
observational: they compare users to non-users and therefore confront selection.
\citet{peet2024infant} pushes the frontier methodologically, using
double/debiased machine learning \citep{chernozhukov2018double} and finding that
doula benefits \emph{grow with predicted infant-health risk}---a targeting result
that motivates my focus on the highest-risk group.

\paragraph{Perinatal policy in health economics.} A literature focused on
identifying the causal effects of perinatal policy on infant health is closely
related. \citet{sonchak2015medicaid} is
the closest analog to this paper: she uses Medicaid reimbursement rates as an
instrument for prenatal care and finds birth-weight gains concentrated among
disadvantaged and Black mothers. \citet{guldi2023effects} find pregnancy-Medicaid
expansions produced modest birth-outcome gains and, notably, reductions in
maternal stress and depression---a candidate mechanism for any maternal-support
policy. \citet{cyganrehm2022effects} show that incentivizing early prenatal care
improves neonatal outcomes. Finally, the birth-attendant-workforce papers of
\citet{anderson2020occupational} and \citet{noghani2023longterm}, exploiting the
staggered adoption of early midwifery laws, find long-run health gains that are
\emph{larger for Black populations}---a close parallel to the present setting,
where a birth-support workforce is expanded by policy and the gains concentrate
among Black mothers.

My contribution sits at the intersection: I bring quasi-experimental
(policy-timing) identification to the doula question, and I find the same heterogeneity logic that
runs through \citet{peet2022variation}, \citet{sonchak2015medicaid}, and
\citet{anderson2020occupational}---benefits concentrated where baseline risk is
highest.

\section{Policy Setting and Data}\label{sec:data}
\subsection{Medicaid doula coverage mandates}
\begin{figure}[H]
\centering
\includegraphics[width=0.49\linewidth]{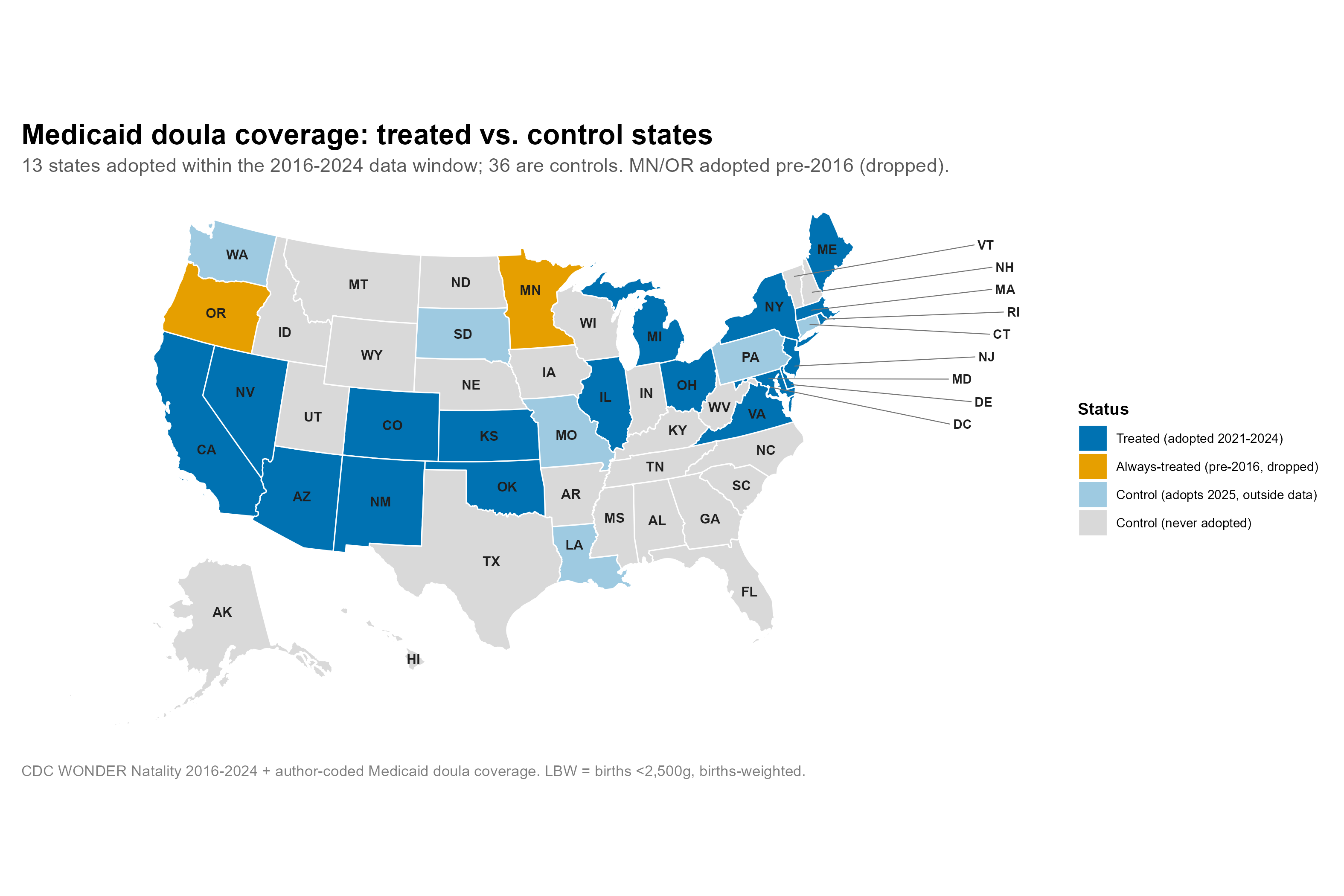}\hfill
\includegraphics[width=0.49\linewidth]{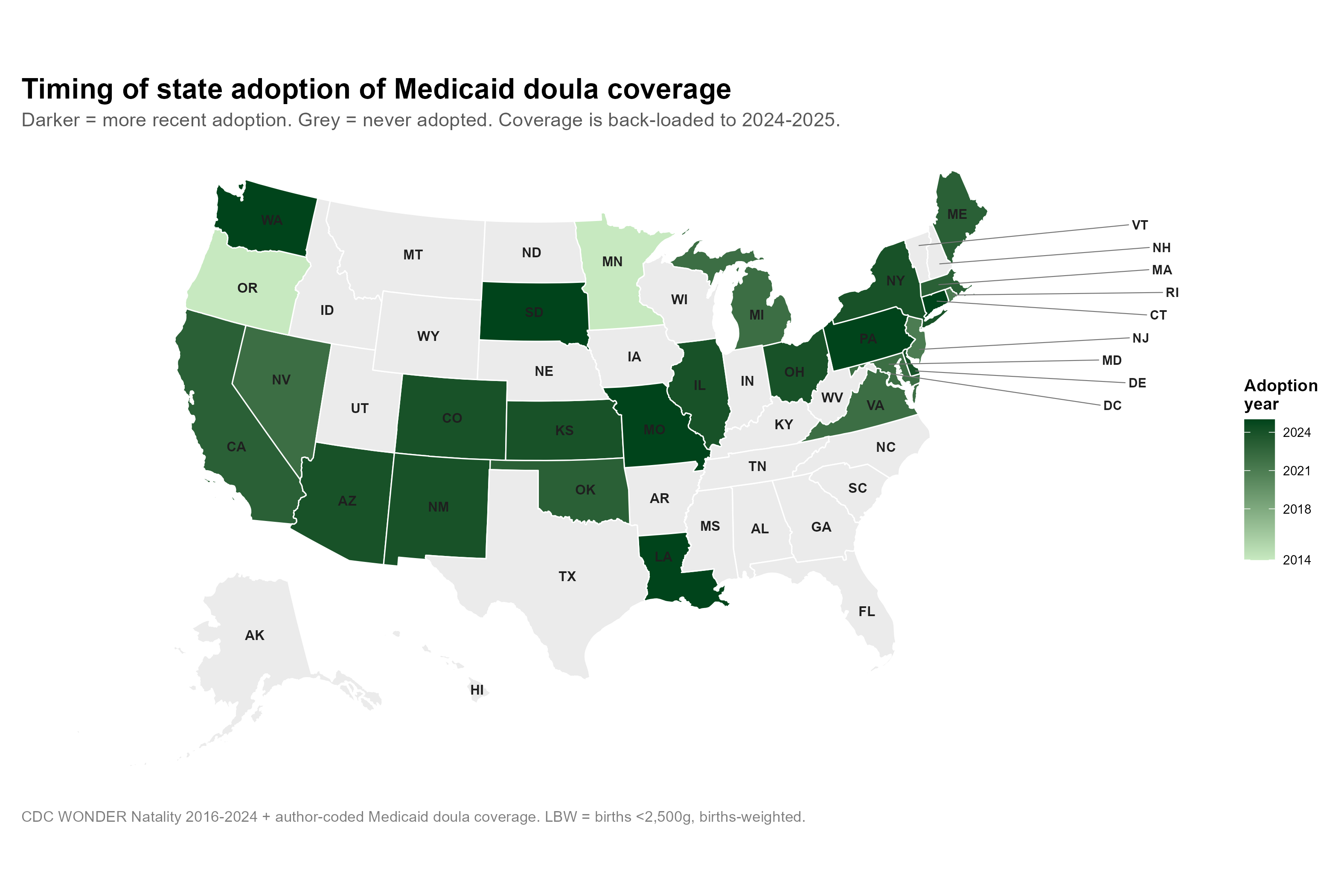}
\caption{Left: treated (adopted within the 2016--2024 window) versus control
states; Minnesota and Oregon (pre-2016 adopters) are dropped. Right: timing of
adoption, lighter (earlier) to darker (more recent). Coverage is recent and
back-loaded.}
\label{fig:maps}
\end{figure}

Minnesota (2014) and Oregon (2014) were the first states to reimburse doula
services through Medicaid. Broad adoption began only in 2021, and accelerated
sharply: by my coding, 13 states began statewide coverage during the 2016--2024
data window, and a further wave took effect in 2025. Figure~\ref{fig:maps}
displays the rollout. Two features matter for what follows. First, adoption is
\emph{geographically selected}: early adopters are concentrated in coastal and
Midwestern states (Figure~\ref{fig:maps}, left), not in the high-burden Deep
South. Second, adoption is \emph{back-loaded}: most coverage takes effect in
2024--2025 (Figure~\ref{fig:maps}, right), at or after the end of the birth data.
Both features---selection and timing---shape identification and power.

\subsection{Birth data and outcomes}
The outcome data are the universe of U.S.\ births from CDC WONDER Natality,
2016--2024, tabulated by state, year, maternal race, maternal education,
birth-weight bin, and tobacco use---32.1 million births in total. The series
begins in 2016 for practical data-access reasons: the CDC WONDER interface limits
the span of years that can be extracted, so I retrieved the most recent available
window. This is not a binding limitation for the question at hand, because
essentially all of the policy variation falls within it---statewide Medicaid doula
coverage is a phenomenon of 2021 onward---so the recent years are precisely the
ones that identify the effect. I aggregate to
the state--year level (and, for heterogeneity, to state--year--group cells),
weighting by births. My primary outcome is the LBW rate, the share of births
under 2{,}500 grams; I also examine very low birth weight (VLBW, $<$1{,}500g) and,
to characterize where in the distribution the policy acts, the share of births in
each of the twelve birth-weight bins. I drop Minnesota and Oregon (always-treated
before the sample), leaving 13 treated and 36 never-treated states, 441
state--year observations. Table~\ref{tab:sumstats} reports summary statistics.

\begin{table}[H]
\centering
\caption{Summary statistics (unit: state--year, $N=441$, 2016--2024)}
\label{tab:sumstats}
\begin{threeparttable}
\begin{tabular}{lcccc}
\toprule
Variable & Mean & SD & Min & Max \\
\midrule
Low birth weight (\%)        & 7.48  & 1.57  & 2.54 & 12.12 \\
Very low birth weight (\%)   & 0.88  & 0.45  & 0.00 & 1.94  \\
Mother's age (years)         & 28.99 & 1.10  & 26.59& 31.86 \\
Gestation (weeks)            & 38.63 & 0.20  & 38.04& 39.24 \\
Prenatal visits (\#)         & 11.18 & 0.68  & 9.70 & 13.30 \\
Smoking (\%)                 & 6.13  & 4.27  & 0.09 & 24.65 \\
Black births (\%)            & 14.18 & 12.07 & 0.00 & 56.91 \\
White births (\%)            & 75.87 & 14.00 & 24.80& 99.34 \\
$\leq$High school (\%)       & 36.90 & 6.22  & 22.62& 49.11 \\
Bachelor's$+$ (\%)           & 34.18 & 7.99  & 19.33& 53.64 \\
Births per state--year       & 72{,}767 & 85{,}524 & 4{,}502 & 487{,}271 \\
\bottomrule
\end{tabular}
\begin{tablenotes}\footnotesize
\item Source: CDC WONDER Natality 2016--2024. Minnesota and Oregon excluded
(always-treated). Rates are births-weighted within state--year. CDC suppresses
cells with fewer than 10 births; the very-low-birth-weight and Black-share minima
of 0.00 therefore reflect small-state suppression rather than true zeros.
\end{tablenotes}
\end{threeparttable}
\end{table}

\subsection{Disparities and geography}
\begin{figure}[H]
\centering
\includegraphics[width=0.49\linewidth]{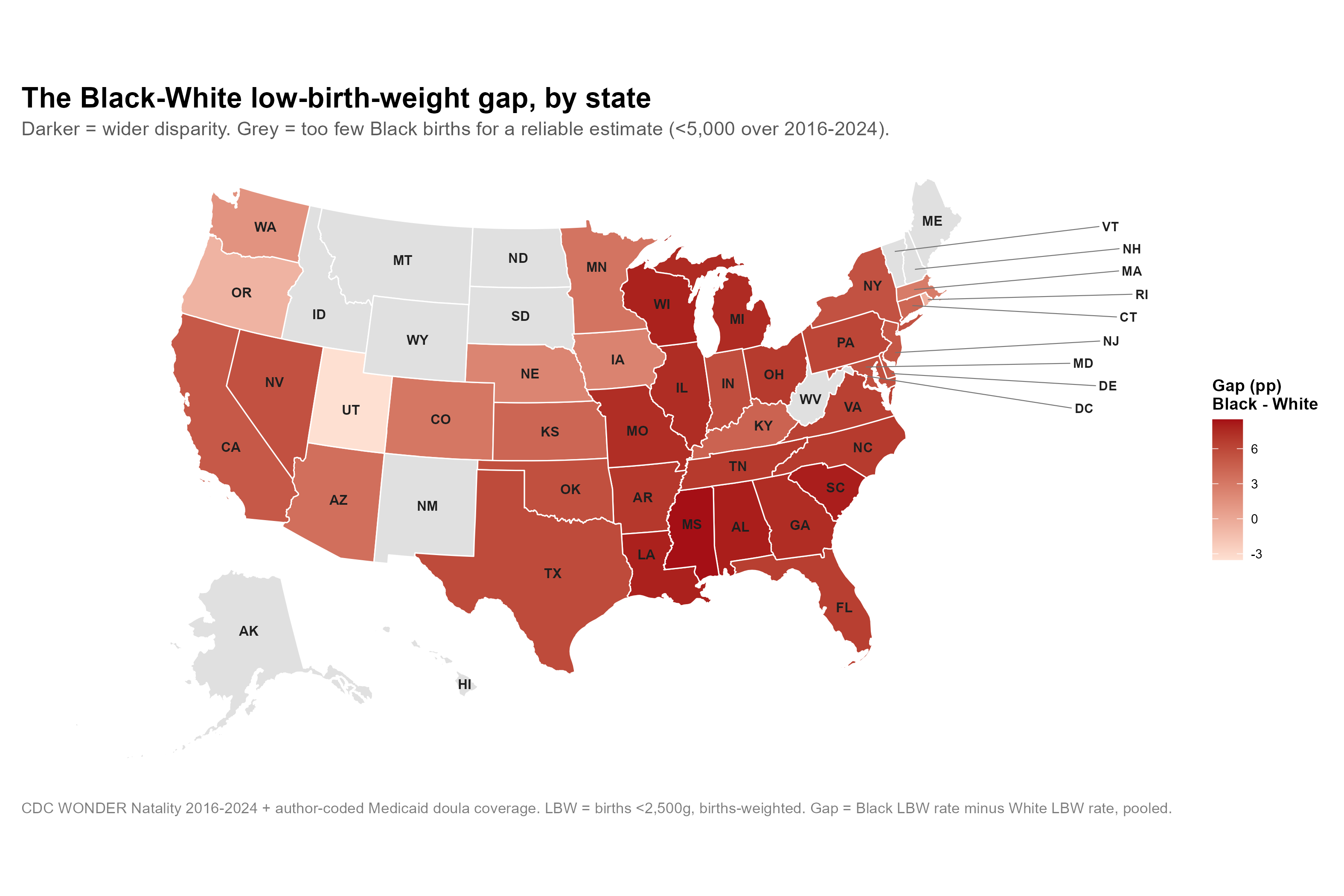}\hfill
\includegraphics[width=0.49\linewidth]{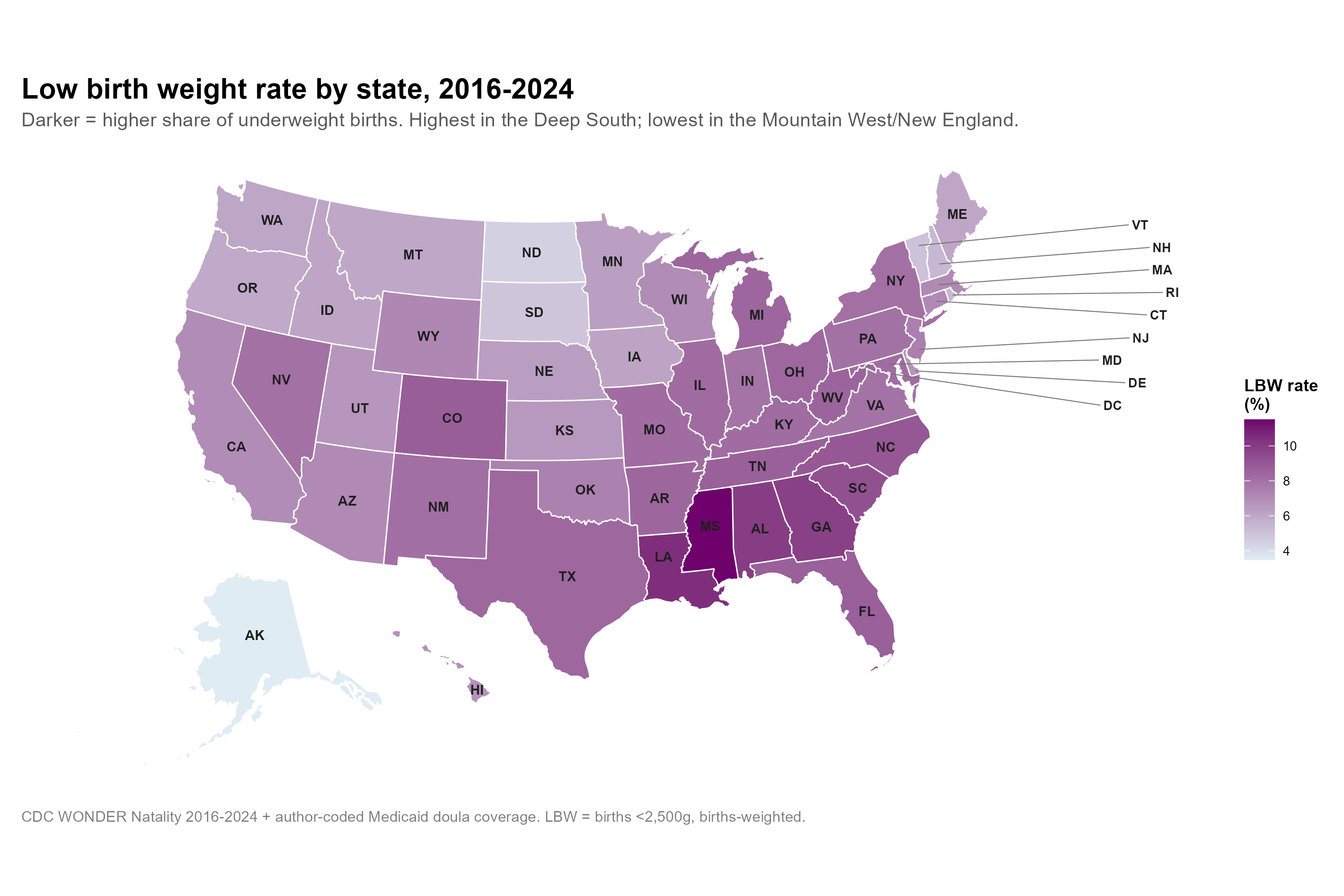}
\caption{Left: Black--white LBW gap by state (grey = too few Black births for a
reliable estimate). Right: overall LBW rate by state. Need is greatest in the
Deep South---largely untreated.}
\label{fig:gapmap}
\end{figure}

Figure~\ref{fig:gap} already documented the central motivating fact: a large,
widening Black--white LBW gap. The disparity is also geographic. The Black--white
gap is widest in the Deep South and industrial Midwest (Figure~\ref{fig:gapmap},
left), and overall LBW is highest in the same Deep South states
(Figure~\ref{fig:gapmap}, right). Comparing these maps to the adoption map
(Figure~\ref{fig:maps}) makes an uncomfortable point plain: the states with the
greatest need are largely the ones that have \emph{not} adopted coverage.

\subsection{Doula supply}
\begin{figure}[H]
\centering
\includegraphics[width=0.72\linewidth]{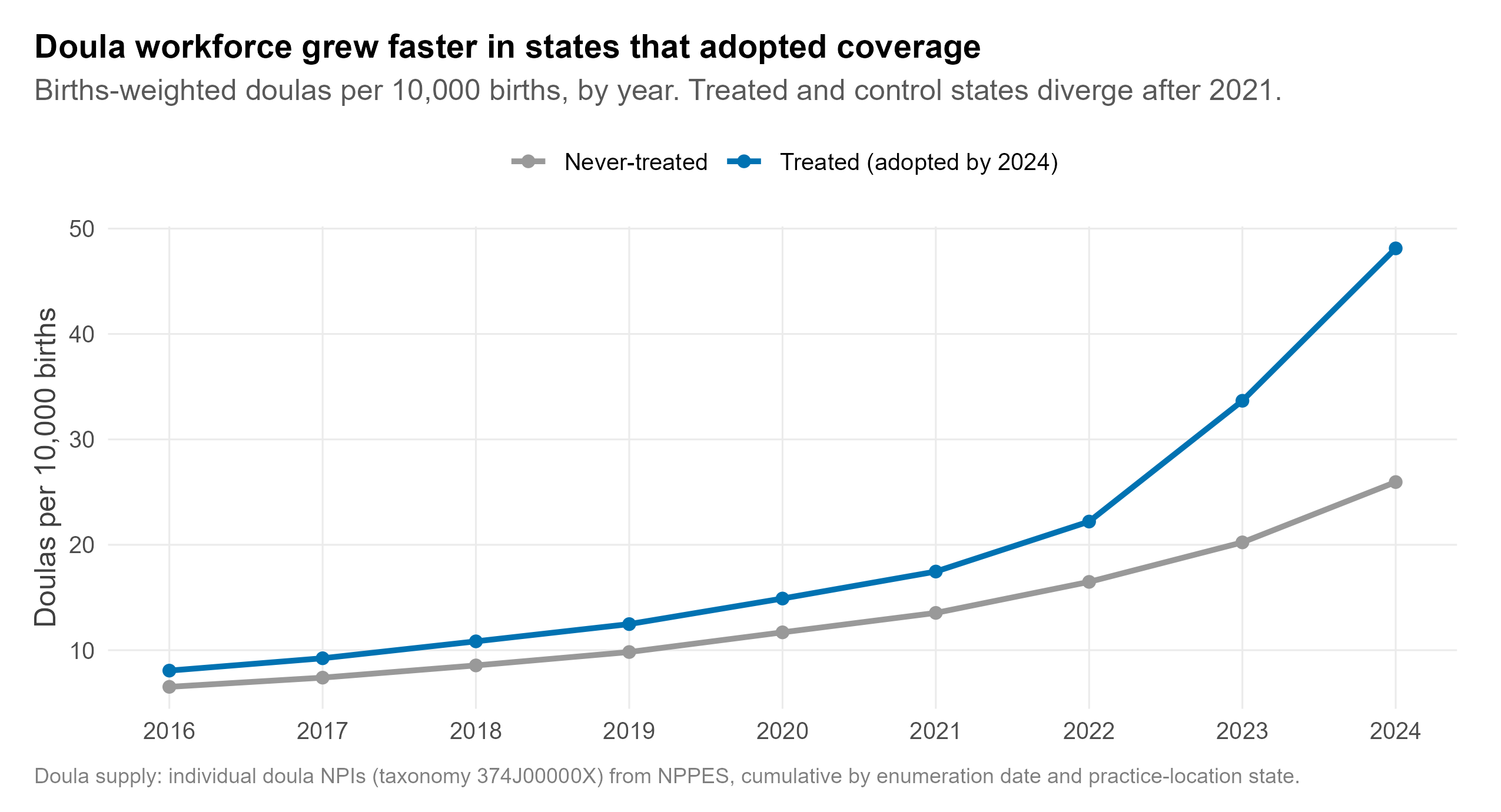}
\caption{Doula supply (births-weighted doulas per 10{,}000 births) by year,
treated versus never-treated states. The series diverge after coverage adoption
begins in 2021.}
\label{fig:doulatrend}
\end{figure}

To study the mechanism, I measure the doula workforce directly. From the National
Plan and Provider Enumeration System (NPPES) I extract every individual provider
with the doula taxonomy code (374J00000X)---19{,}425 providers---together with
their practice-location state and enumeration year, and construct the cumulative
stock of doulas per 10{,}000 births by state and year. By 2024, treated states
average about 51 doulas per 10{,}000 births versus 26 in control states.
Figure~\ref{fig:doulatrend} plots the series: the two groups track closely through
2020 and diverge after coverage begins in 2021. This is a measure of registered
doula supply, not of Medicaid-billed services; I return to its limitations in
Section~\ref{sec:discussion}.

\section{Empirical Strategy}\label{sec:strategy}
\subsection{Difference-in-differences}
Let $Y_{st}$ denote a birth outcome (e.g.\ the LBW rate) in state $s$ and year
$t$. The baseline TWFE specification is
\begin{equation}
Y_{st} = \beta\, D_{st} + \alpha_s + \lambda_t + \varepsilon_{st},
\label{eq:twfe}
\end{equation}
where $D_{st}$ indicates that state $s$ has active coverage in year $t$, $\alpha_s$
and $\lambda_t$ are state and year fixed effects, observations are weighted by
births, and standard errors are clustered by state. A state is coded as treated
in the first year for which births could plausibly be exposed (coverage effective
on or before mid-year).

Because adoption is staggered and effects may be heterogeneous across cohorts,
the TWFE estimator in \eqref{eq:twfe} can be biased
\citep{goodmanbacon2021difference,dechaisemartin2020twoway}. I therefore report
the group-time estimator of \citet{callaway2021difference} (doubly robust,
not-yet-treated comparison group, multiplier-bootstrap inference) as the primary
specification for dynamics and for the distributional analysis, and I show
event-study estimates to assess parallel trends.

\subsection{Identification and its threats}
Identification requires that, absent coverage, treated and control states would
have followed parallel LBW paths. Three threats deserve emphasis.

\emph{Selection.} Treated states are positively selected on socioeconomic status.
Table~\ref{tab:balance} compares pre-period (2016--2020) means: treated states
have older, more educated, less-smoking mothers, more prenatal care, and more
generous Medicaid. Reassuringly, the LBW \emph{level} itself is balanced
($p=0.45$), so treated and control states start from similar outcomes; but the
covariate gaps mean raw cross-state comparisons do not identify causal effects,
and motivate the within-state, policy-timing approach.

\emph{Staggered timing.} Addressed by the \citet{callaway2021difference}
estimator, as above.

\emph{Power.} With most adoption back-loaded to 2024--2025, treated person-time in
the data is small. This limits precision throughout and is the central caveat of
the paper.

\begin{table}[H]
\centering
\caption{Pre-period balance (2016--2020): treated vs.\ never-treated states}
\label{tab:balance}
\begin{threeparttable}
\begin{tabular}{lcccc}
\toprule
Variable & Treated & Never-treated & Difference & $p$-value \\
\midrule
Low birth weight (\%)      & 7.26 & 7.39 & $-0.13$ & 0.452 \\
Very low birth weight (\%) & 0.90 & 0.88 & $+0.02$ & 0.671 \\
Mother's age (years)       & 29.46& 28.48& $+0.98$ & 0.000 \\
Gestation (weeks)          & 38.72& 38.65& $+0.08$ & 0.002 \\
Prenatal visits (\#)       & 11.51& 11.21& $+0.30$ & 0.005 \\
Smoking (\%)               & 5.68 & 8.63 & $-2.95$ & 0.000 \\
Black births (\%)          & 16.40& 13.74& $+2.65$ & 0.069 \\
$\leq$High school (\%)     & 35.93& 37.68& $-1.75$ & 0.051 \\
Bachelor's$+$ (\%)         & 35.83& 31.66& $+4.18$ & 0.000 \\
Medicaid threshold (\% FPL)& 208.8& 195.7& $+13.1$ & 0.226 \\
\bottomrule
\end{tabular}
\begin{tablenotes}\footnotesize
\item Means across pre-period state--years. The LBW level is balanced; treated
states are positively selected on SES.
\end{tablenotes}
\end{threeparttable}
\end{table}


\subsection{Mechanism: coverage as an instrument for doula supply}
To connect the policy to doula activity I estimate a two-stage least squares
(instrumental-variables) model in which doula supply $S_{st}$ (doulas per 10{,}000 births) is
the endogenous regressor and coverage $D_{st}$ is the instrument:
\begin{equation}
S_{st} = \pi D_{st} + \alpha_s + \lambda_t + u_{st},\qquad
Y_{st} = \theta\, \widehat{S}_{st} + \alpha_s + \lambda_t + \nu_{st}.
\label{eq:iv}
\end{equation}
The exclusion restriction is that mandates affect Black LBW only through doula
supply; I discuss its plausibility in Section~\ref{sec:discussion}.

\section{Results}\label{sec:results}
\subsection{The average effect is a precise null}
\begin{figure}[H]
\centering
\includegraphics[width=0.95\linewidth]{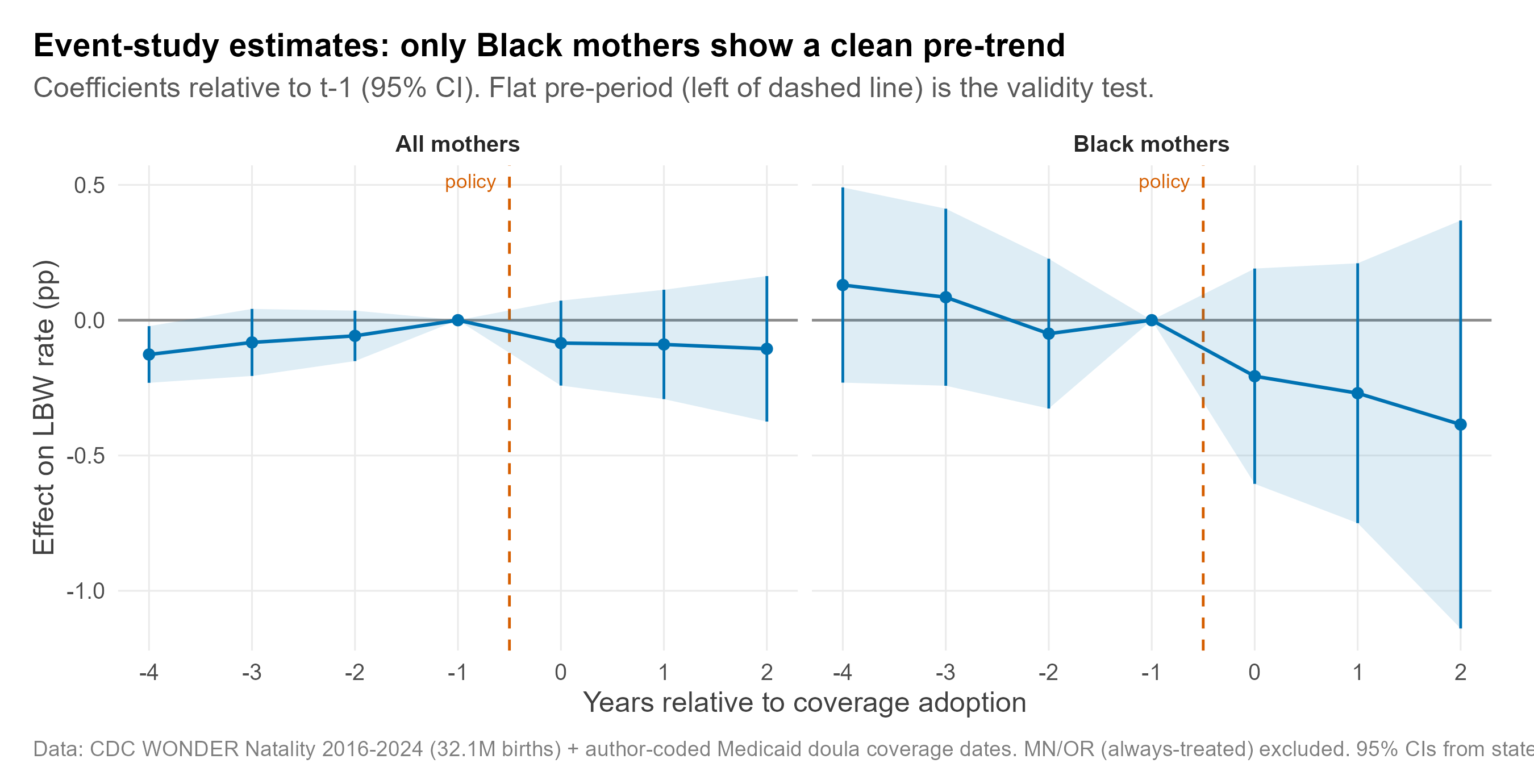}
\caption{Event-study estimates relative to the year before adoption (95\% CI).
For all mothers (left) the path is flat throughout. For Black mothers (right) the
pre-period is flat---supporting parallel trends---and the outcome declines after
adoption.}
\label{fig:event}
\end{figure}

I begin with the average effect, briefly, because it motivates everything that
follows. The TWFE estimate from \eqref{eq:twfe} is $-0.010$ percentage points
($p=0.92$): a precise zero. The event study (Figure~\ref{fig:event}, left panel)
tells the same story for all mothers---no break at adoption. Read on its own, this
would be a null result. But, exactly as \citet{peet2022variation} warn, an
average effect on an infrequent outcome can hide large benefits for the highest-risk
group. The rest of the paper asks whether it does.

\subsection{Heterogeneity: the effect concentrates among Black mothers}
\begin{figure}[H]
\centering
\includegraphics[width=0.95\linewidth]{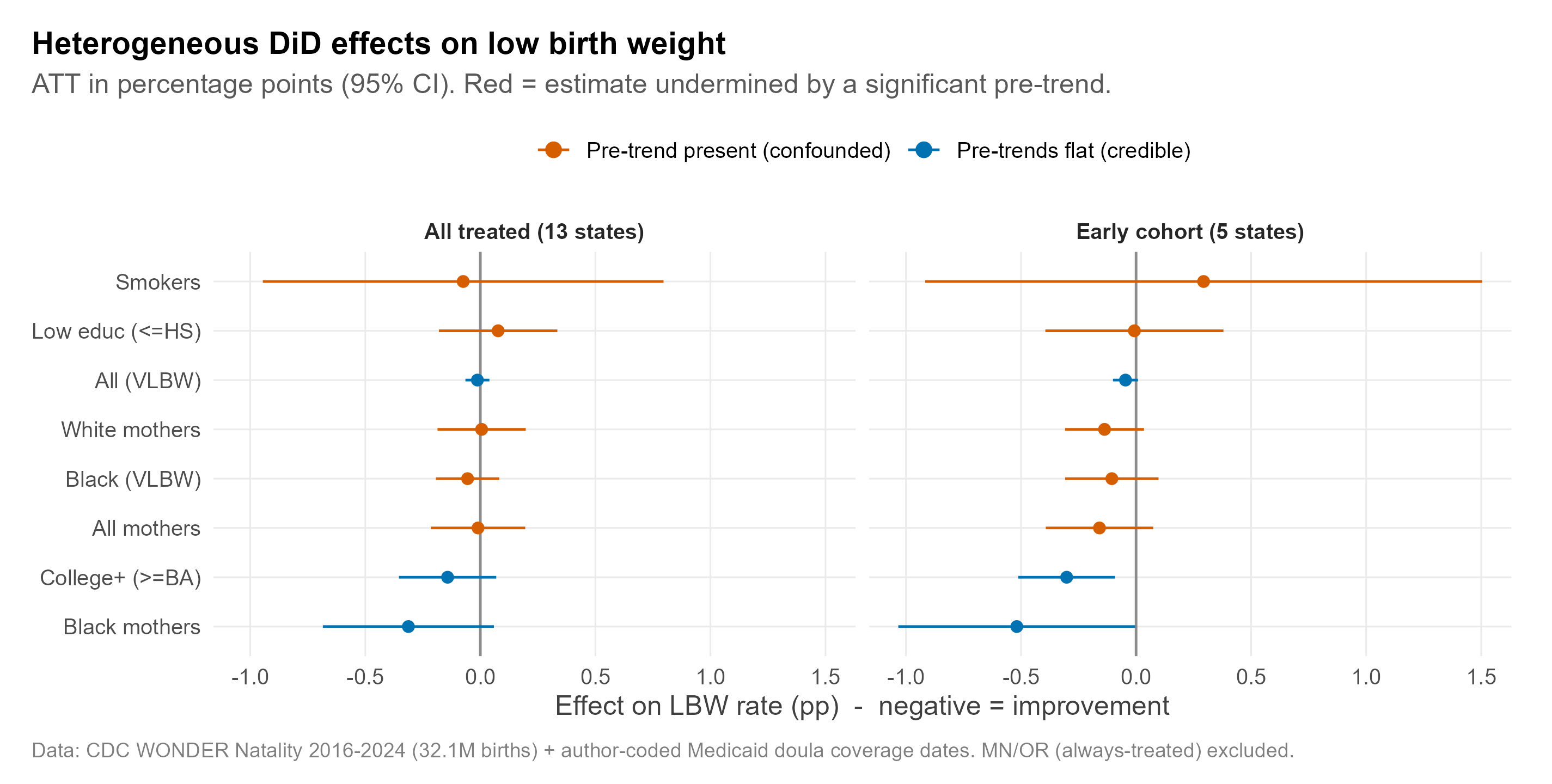}
\caption{Two-way fixed-effects DiD effects on LBW by subgroup (early-cohort ATT,
95\% CI). Red points are undermined by a significant pre-trend; blue points have
flat pre-trends. Black mothers are the one credible negative estimate. The
corresponding Callaway--Sant'Anna event-study graphs for every subgroup are in
Appendix Figure~\ref{fig:cssub}.}
\label{fig:hetero}
\end{figure}

The heart of the paper is heterogeneity. Figure~\ref{fig:hetero} reports DiD
estimates by subgroup. Two things stand out. First, several subgroups show
negative point estimates, but for all mothers, white mothers, lower-education
mothers, and smokers these are accompanied by \emph{significant pre-trends}
(plotted in red): their apparent ``improvement'' is a pre-existing trend, not a
treatment effect. Second, Black mothers are the exception: a sizable negative
estimate \emph{with flat pre-trends}. Concretely, the pre-trend test---the
$t{-}4$ event-study coefficient relative to $t{-}1$---is insignificant only for
Black mothers ($p=0.48$); for all mothers, white mothers, lower-education mothers,
and smokers it is significant (Table~\ref{tab:hetero}, final column), so the
``effect'' for those groups is a continuation of a pre-existing trend. Black
mothers are thus the only LBW subgroup whose estimate is both economically
meaningful and free of a pre-trend---and it is the subgroup that theory
\citep{peet2022variation,sonchak2015medicaid,anderson2020occupational} and the
disparity facts of Section~\ref{sec:data} single out in advance.

Table~\ref{tab:hetero} reports the underlying estimates. The pattern is clear:
Black mothers are the only LBW subgroup with both an economically meaningful
negative estimate and flat pre-trends.

One other estimate is statistically notable and deserves comment: the
early-cohort effect for college-educated (Bachelor's$+$) mothers is sizable
($-0.302$) and conventionally significant, with flat pre-trends ($p=0.90$). I do
not interpret it as a doula effect. College-educated mothers are largely above
Medicaid income thresholds and so are mostly ineligible for the benefit, which
makes a true causal effect of doula \emph{coverage} on this group implausible; a
mechanical channel is absent. The most likely explanations are that the estimate
reflects something correlated with the timing of adoption in these states (for
example, other contemporaneous policies) or that it is a false positive among the
several subgroups examined---under a multiple-comparison correction it would not
survive. It is, in other words, a useful cautionary contrast: a ``significant''
result can arise where no mechanism plausibly operates, which is precisely why I
lean on theory and pre-trends, not stars alone, to identify the credible estimate.

\begin{table}[H]
\centering
\caption{Difference-in-differences estimates on low birth weight, by subgroup}
\label{tab:hetero}
\begin{threeparttable}
\begin{tabular}{lccc}
\toprule
 & ATT, all-treated & ATT, early cohort & Pre-trend \\
Subgroup & (13 states) & (5 states) & test $p$ \\
\midrule
\multicolumn{4}{l}{\emph{Panel A. Low birth weight ($<$2{,}500g)}}\\
All mothers       & $-0.010$        & $-0.159$         & 0.018 \\
                  & (0.105)         & (0.119)          &       \\
Black mothers     & $-0.313^{*}$    & $-0.519^{**}$    & \textbf{0.480} \\
                  & (0.190)         & (0.263)          &       \\
White mothers     & $0.006$         & $-0.137$         & 0.002 \\
                  & (0.098)         & (0.087)          &       \\
$\leq$High school & $0.077$         & $-0.007$         & 0.008 \\
                  & (0.131)         & (0.197)          &       \\
Bachelor's$+$     & $-0.142$        & $-0.302^{***}$   & 0.898 \\
                  & (0.108)         & (0.107)          &       \\
Smokers           & $-0.075$        & $0.293$          & 0.018 \\
                  & (0.444)         & (0.618)          &       \\
\addlinespace
\multicolumn{4}{l}{\emph{Panel B. Very low birth weight ($<$1{,}500g)}}\\
All mothers       & $-0.013$        & $-0.046^{*}$     & 0.303 \\
                  & (0.027)         & (0.028)          &       \\
Black mothers     & $-0.055$        & $-0.105$         & 0.078 \\
                  & (0.070)         & (0.104)          &       \\
\bottomrule
\end{tabular}
\begin{tablenotes}\footnotesize
\item ATT in percentage points; births-weighted TWFE with state and year fixed
effects; state-clustered SE in parentheses. $^{*}p<0.10$, $^{**}p<0.05$,
$^{***}p<0.01$ (two-sided, naive). The final column reports the $p$-value of the
pre-trend test---the $t{-}4$ event-study coefficient relative to $t{-}1$---so that a
\emph{large} $p$ indicates no evidence against parallel trends. Among LBW
subgroups, only Black mothers have a large pre-trend $p$ (0.48); for all mothers,
white mothers, lower-education mothers, and smokers the pre-trend is significant,
so their negative estimates reflect pre-existing trends. \emph{Caution:} stars use
naive cluster-robust SEs; valid few-cluster inference for the Black early-cohort
estimate (Table~\ref{tab:inference}) places its two-sided $p\approx0.10$. The
Bachelor's$+$ estimate, though starred, is theoretically implausible
(college-educated mothers are rarely Medicaid-eligible) and illustrates the
multiple-testing risk.
\end{tablenotes}
\end{threeparttable}
\end{table}

\begin{figure}[H]
\centering
\includegraphics[width=0.49\linewidth]{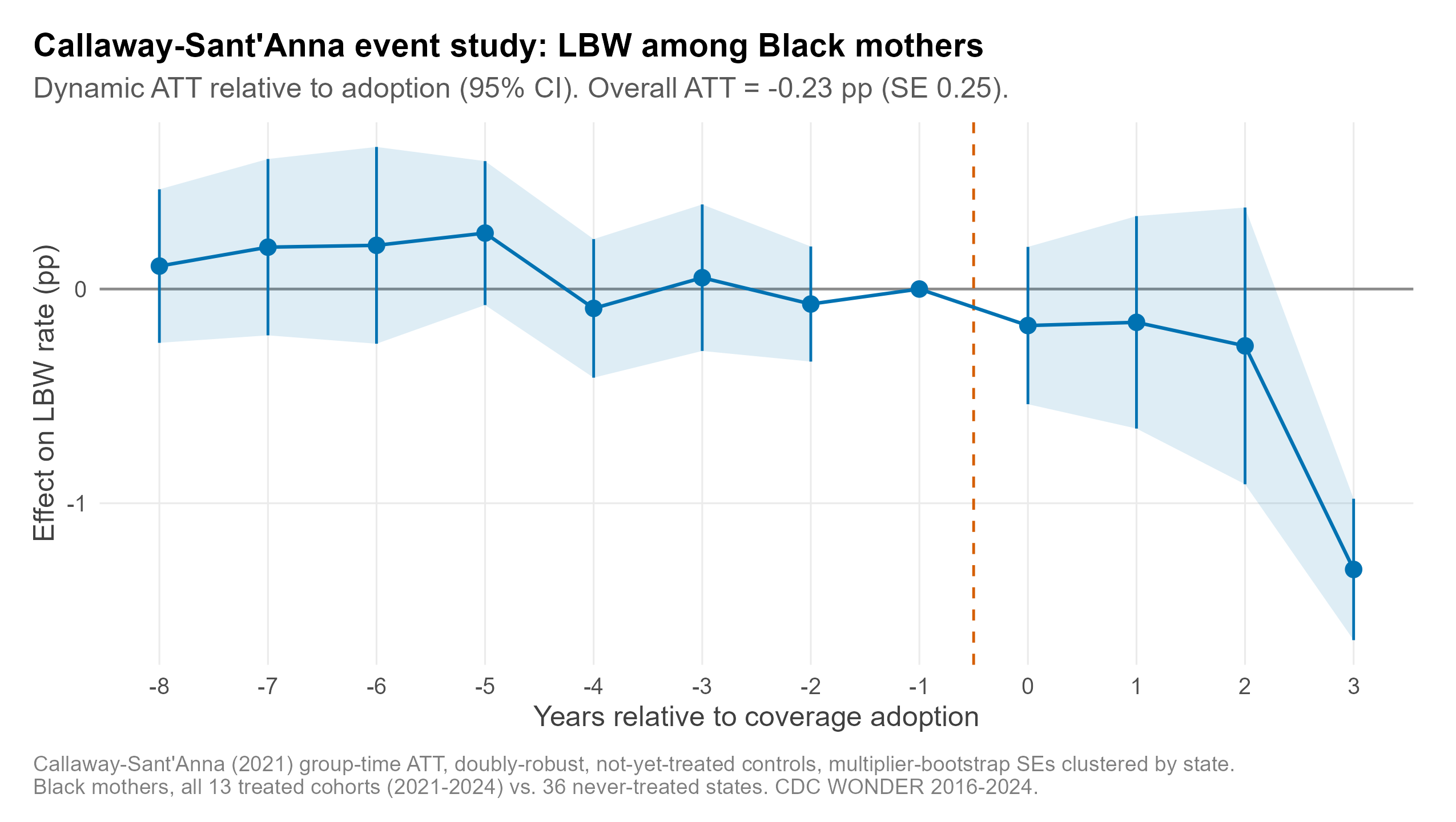}\hfill
\includegraphics[width=0.49\linewidth]{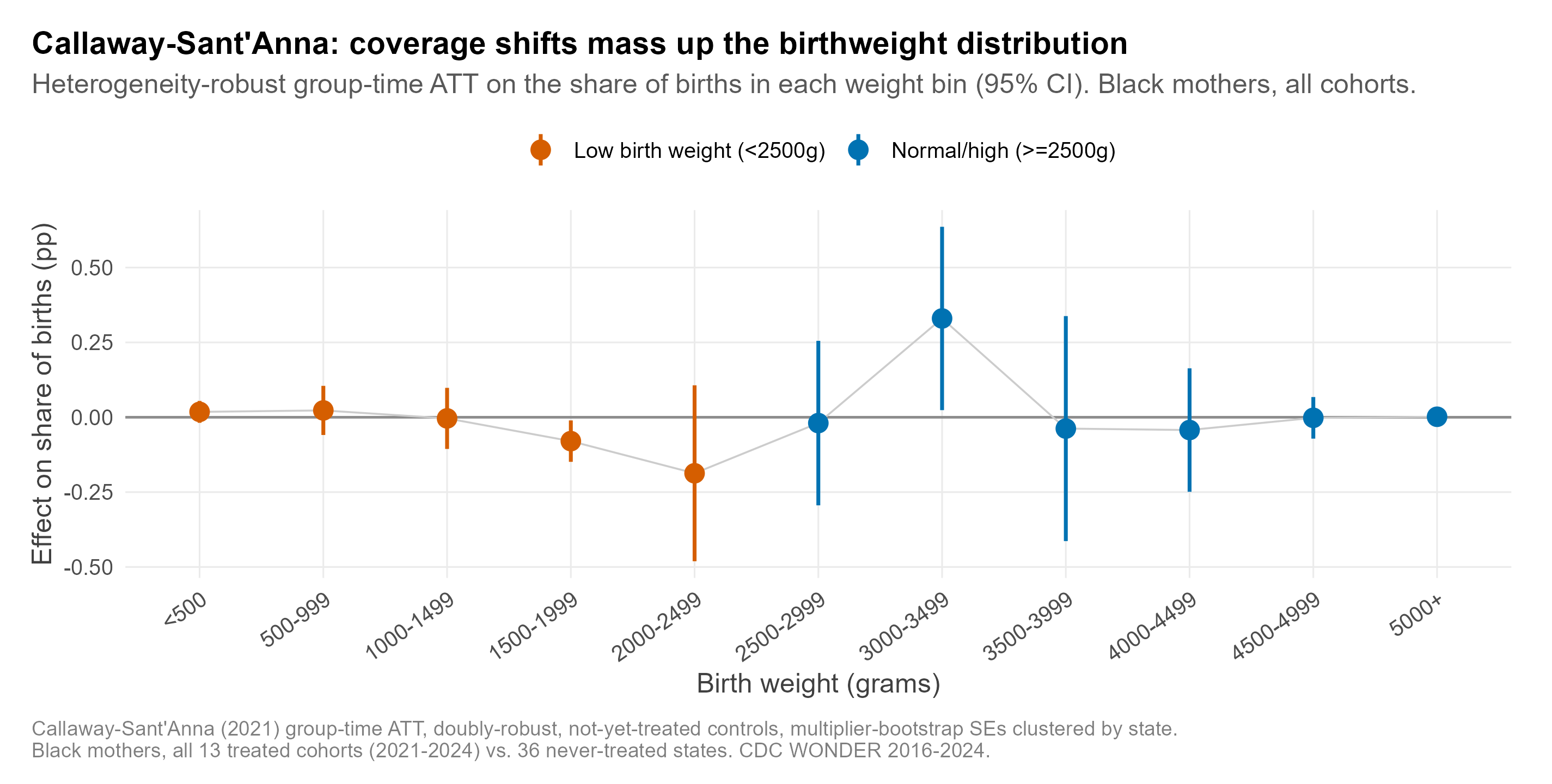}
\caption{Callaway--Sant'Anna estimates for Black mothers. Left: dynamic effect on
LBW---flat pre-trends, growing with exposure. Right: effect across the
birth-weight distribution---mass shifts out of the low-weight bins and into the
normal range.}
\label{fig:cs}
\end{figure}

For Black mothers in the earliest-adopting cohort (New Jersey, Maryland, Nevada,
Rhode Island, Virginia---the five states with at least two post-period years), LBW
falls by $0.52$ percentage points on a $9.9\%$ base, about a $5\%$ relative
reduction. The Callaway--Sant'Anna event study (Figure~\ref{fig:cs}, left)
confirms flat pre-trends and an effect that \emph{grows with exposure}: near zero
in the first post-year and larger thereafter---precisely the dynamic one expects
from a workforce that takes time to scale and a benefit that accrues over a
pregnancy. Pooling all cohorts, including the barely-exposed 2023--2024 adopters,
the Callaway--Sant'Anna average is $-0.23$ percentage points and statistically
insignificant: the recent adopters dilute the average toward zero, which is why
the effect is visible only where exposure is long.

A distributional view sharpens the interpretation. Rather than relying on the
single 2{,}500g threshold, Figure~\ref{fig:cs} (right) estimates the effect on the
share of births in each weight bin. The pattern is a coherent \emph{upward shift}:
a significant decline in the share of births in low-weight bins
(e.g.\ 1{,}500--1{,}999g, $-0.08$pp) and a significant increase in the
normal-weight range (3{,}000--3{,}499g, $+0.33$pp), with the curve crossing zero
near the clinical threshold. Such a shape is hard to generate by chance and is
more informative than any single cutoff; Appendix Table~\ref{tab:csbins} reports
the full set of bin-by-bin estimates.

\subsection{Inference with few treated clusters}
\begin{figure}[H]
\centering
\includegraphics[width=0.75\linewidth]{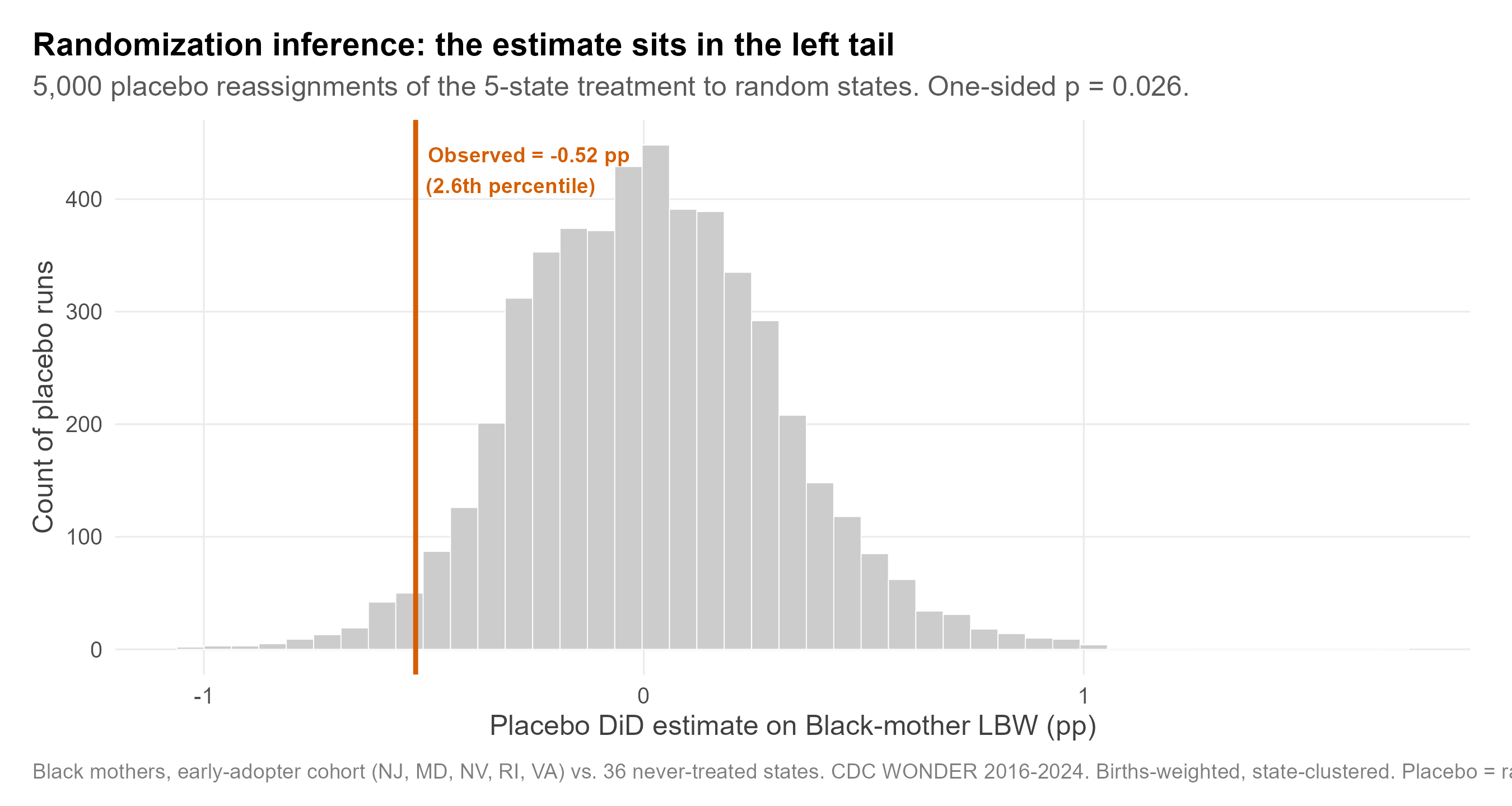}
\caption{Randomization inference: the observed Black-mother estimate (red line)
lies in the left tail of 5{,}000 placebo estimates from random treatment
assignments.}
\label{fig:ri}
\end{figure}

Because the Black-mother result rests on only five treated states, conventional
cluster-robust standard errors are unreliable \citep{cameron2008bootstrap,
mackinnon2017wild}. I therefore stress-test the early-cohort estimate three ways
(Table~\ref{tab:inference}). The naive cluster-robust $p$-value is about $0.05$;
the wild cluster bootstrap (Rademacher and Webb weights) and randomization
inference (5{,}000 placebo reassignments) all place the two-sided $p$-value near
$0.10$, with the observed estimate in the left tail of the placebo distribution
(Figure~\ref{fig:ri}). The honest reading is that the result is \emph{marginal}:
significant one-sided ($p\approx0.03$) but not at conventional two-sided levels.
Two further checks support it: the magnitude is stable when any single treated
state is dropped (it is not driven by one state; Appendix
Figure~\ref{fig:channels}, right panel), and it survives a one-year gestational
lag in treatment timing---the early-cohort estimate moves only from $-0.52$ to
$-0.41$ pp---albeit with reduced precision as the lag shortens the already-thin
post-period.

\begin{table}[H]
\centering
\caption{Inference for the Black-mother early-cohort estimate (ATT $=-0.52$ pp)}
\label{tab:inference}
\begin{threeparttable}
\begin{tabular}{lcc}
\toprule
Inference method & Two-sided $p$ & \\
\midrule
Naive cluster-robust SE                       & 0.05 & \\
Wild cluster bootstrap (Rademacher, $B=9{,}999$) & 0.102 & \\
Wild cluster bootstrap (Webb 6-point, $B=9{,}999$) & 0.103 & \\
Randomization inference (5{,}000 placebos)     & 0.102 & \\
\midrule
One-sided directional test (RI)                & 0.03 & \\
\bottomrule
\end{tabular}
\begin{tablenotes}\footnotesize
\item Five treated states (NJ, MD, NV, RI, VA) vs.\ 36 never-treated.
Births-weighted, state-clustered.
\end{tablenotes}
\end{threeparttable}
\end{table}

\subsection{Mechanism: coverage expands the doula workforce}
\begin{figure}[H]
\centering
\includegraphics[width=0.49\linewidth]{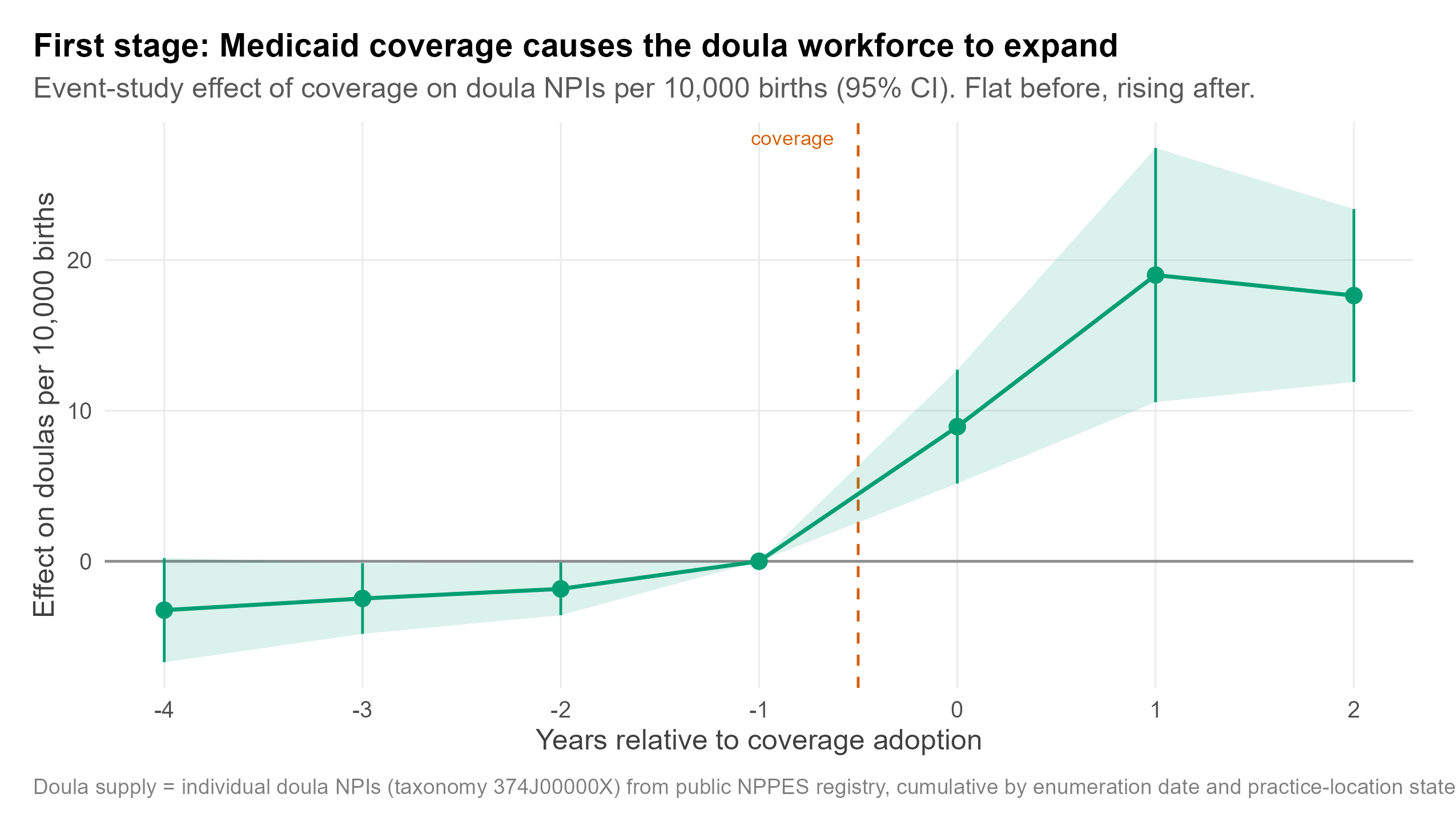}\hfill
\includegraphics[width=0.49\linewidth]{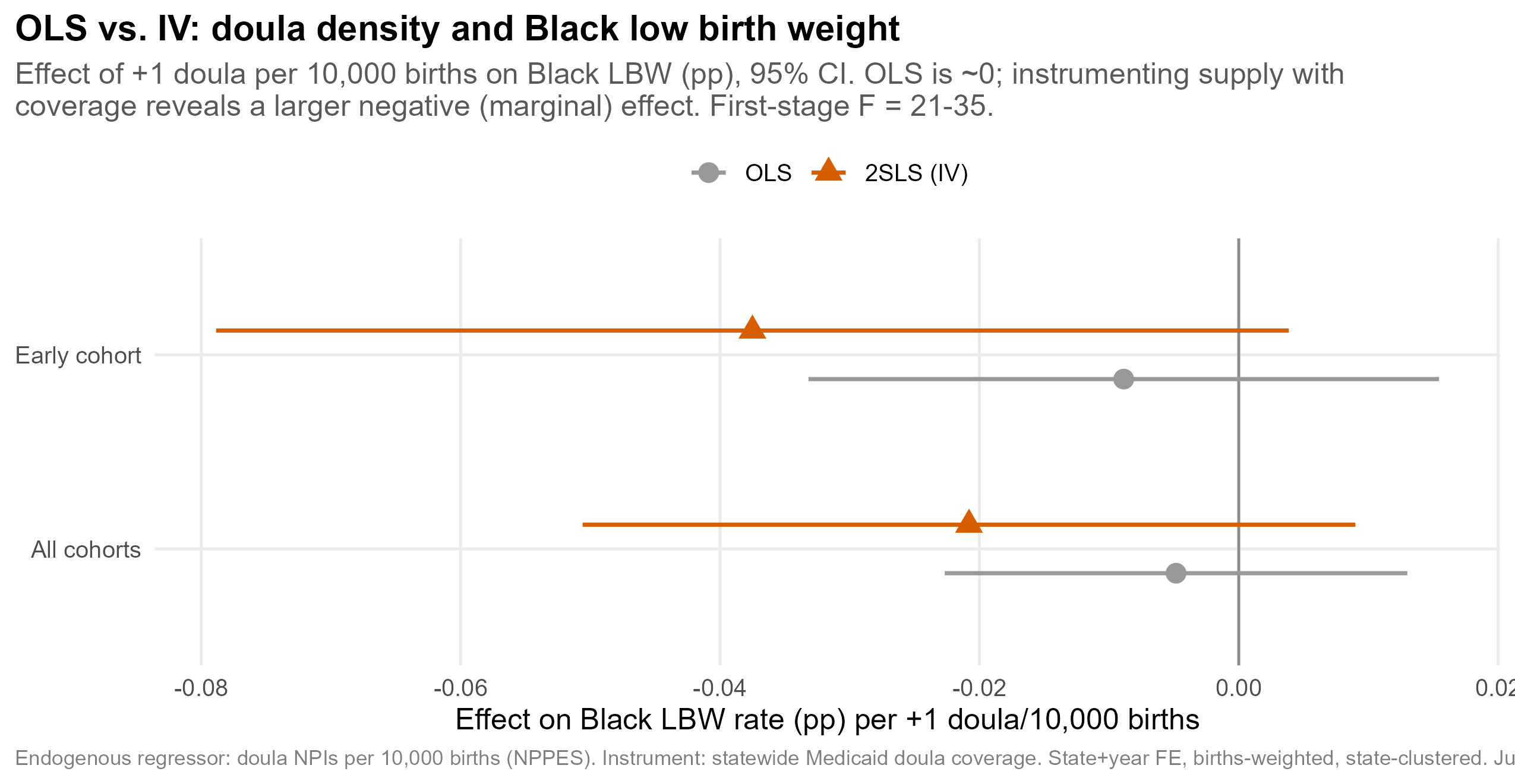}
\caption{Mechanism. Left: first stage---coverage expands the doula workforce.
Right: OLS vs.\ 2SLS---instrumenting doula supply with coverage reveals a larger
(marginal) negative effect on Black LBW.}
\label{fig:mech}
\end{figure}

Finally, the mechanism. The first stage is strong and clean: coverage raises the
doula workforce by about 15 doulas per 10{,}000 births
(Figure~\ref{fig:mech}, left), an effect that is flat before adoption and rises
after, with a first-stage $F$ of 20.8 in the all-cohort sample and 35.0 in the
early-cohort sample (Table~\ref{tab:iv})---in both cases comfortably above
conventional weak-instrument thresholds. The two-stage least squares estimate of the effect of doula supply on
Black LBW (Figure~\ref{fig:mech}, right) is negative and larger in magnitude than
the corresponding OLS association, which is near zero---the standard pattern when
an instrument isolates policy-driven variation. The estimate is marginally
significant in the early cohort ($-0.038$ pp per additional doula per 10{,}000
births, $p=0.075$). Because the design is just-identified, the pieces are
internally consistent: the first-stage increase of about 15 doulas per 10{,}000
births times the two-stage least squares coefficient ($-0.038$) implies
$\approx -0.56$ pp, matching the reduced-form Black-mother estimate of $-0.52$ pp.
I read these estimates as a mechanism check rather than a primary causal estimate,
for two reasons developed in Section~\ref{sec:discussion}: part of the measured
increase in doula supply may reflect existing providers registering to bill the
new benefit rather than entirely new entrants---NPPES records when a provider first
obtains a national identifier, which need not coincide with entry into doula
practice, so the data cannot fully separate the two---and the exclusion
restriction---that mandates
affect Black LBW only through doula supply---is an untestable assumption that other
coincident policies or medical advances could violate.

Table~\ref{tab:iv} reports the estimates. The first stage is strong (Panel A); the
2SLS effect of doula supply on Black LBW is negative and larger than OLS, though
only marginally significant in the early cohort (Panel B).

\begin{table}[H]
\centering
\caption{Mechanism: coverage, doula supply, and Black low birth weight}
\label{tab:iv}
\begin{threeparttable}
\begin{tabular}{lcc}
\toprule
 & All cohorts & Early cohort \\
\midrule
\multicolumn{3}{l}{\emph{Panel A. First stage: coverage $\rightarrow$ doula supply}}\\
Coverage mandate & \multicolumn{2}{c}{$+15.07^{***}\;(3.01)$} \\
First-stage $F$  & 20.8 & 35.0 \\
\addlinespace
\multicolumn{3}{l}{\emph{Panel B. Doula supply $\rightarrow$ Black LBW}}\\
OLS              & $-0.005$  & $-0.009$ \\
                 & (0.009)   & (0.012)  \\
2SLS (IV)        & $-0.021$  & $-0.038^{*}$ \\
                 & (0.015)   & (0.021)  \\
\bottomrule
\end{tabular}
\begin{tablenotes}\footnotesize
\item Panel A: effect of coverage on doulas per 10{,}000 births. Panel B: effect of
doula supply (per 10{,}000 births) on the Black LBW rate, in percentage points per
additional doula; the instrument is the coverage mandate. All specifications
include state and year fixed effects, are births-weighted, and cluster SEs by
state. $^{*}p<0.10$, $^{***}p<0.01$. Just-identified 2SLS.
\end{tablenotes}
\end{threeparttable}
\end{table}

\paragraph{In short.} Where doula coverage has had time to operate, low birth
weight among Black mothers---the group at greatest risk---fell; the average effect
is null because most mandates are too recent to have bitten.

\section{Discussion and Limitations}\label{sec:discussion}
The results should be read as credible early evidence, not as a settled causal
claim, and several limitations bound their interpretation. The dominant one is
exposure. Fourteen of twenty-seven adopting states began coverage in 2024--2025,
so the design has little treated person-time and limited power; no estimator can
manufacture variation that is not yet in the data, and the most likely explanation
for the null average effect is simply that birth-weight improvements from doula
services require a longer time horizon to materialize than the data currently
allow. The natural remedy is replication as 2025--2026 natality becomes available.
A related concern is that the Black-mother result rests on only five treated
states; I have been deliberate about inference valid for few clusters, and the
honest verdict is that the estimate is marginal rather than conventionally
significant.

The instrumental-variables analysis carries two further caveats and should be read
as a mechanism check rather than as the paper's primary causal estimate. First,
the doula-supply measure is imperfect: built from the NPPES registry, it captures
registered doulas rather than Medicaid-billed services, misses informal and
community doulas, and partly reflects registration induced by the policy itself.
The first stage is therefore best read as ``coverage brings doulas into the
billing system,'' which is mechanism-relevant but not the same as creating supply
from scratch. Second, and more fundamentally, the two-stage least squares estimate
relies on an exclusion restriction---that mandates affect Black low birth weight
only through doula supply---that the data cannot verify. The relevant threats are
other determinants of birth outcomes that changed around the same time as doula
adoption: contemporaneous Medicaid eligibility expansions (for example,
ACA-related expansions), other state maternal-health initiatives, or advances in
prenatal and neonatal care. State and year fixed effects absorb common national
shocks and fixed differences across states, but not state-specific policies
coincident with the timing of doula adoption, so the restriction remains an
assumption. It bears emphasis that the absence of movement in prenatal visits,
gestational age, and smoking (Appendix Figure~\ref{fig:channels}) does \emph{not}
support this restriction: those are outcomes doulas might plausibly have
influenced, so their flatness is, if anything, a puzzle for the mechanism rather
than evidence about competing channels.

Two final caveats concern interpretation and scope. Because early adopters are
positively selected and are largely not the high-burden Deep South states
(Figures~\ref{fig:maps}--\ref{fig:gapmap}), extrapolating any effect to the states
where the need is greatest requires caution. And the present design, while it uses
policy variation rather than individual take-up, remains observational rather than
experimental; its credibility rests on the parallel-trends assumption, which the
event-study evidence supports for Black mothers but not for every subgroup.

These limitations map directly to a research agenda: more post-period data,
including the restricted natality and linked infant-mortality files; a power
calculation to characterize the minimum detectable effect of the current design; a
continuous policy-intensity measure such as reimbursement generosity, in the
spirit of \citet{sonchak2015medicaid}; separating new doula entry from
registration of existing providers; and an honest-DiD sensitivity analysis
\citep{rambachan2023more} once pre-trends can be estimated more precisely.

\section{Conclusion}\label{sec:conclusion}
Medicaid doula coverage mandates spread rapidly across the United States in the
hope of improving infant health and narrowing racial disparities. Using the
staggered rollout and 32.1 million births, I find no average effect on low birth
weight, but a marginal, theory-consistent reduction concentrated among Black
mothers---the group at greatest risk and the focus motivated in advance by the
disparity facts and by prior work emphasizing risk-targeted benefits
\citep{peet2022variation,sonchak2015medicaid,anderson2020occupational}. The
effect appears only where coverage has operated longest, grows with exposure,
shows up as a coherent upward shift in the birth-weight distribution, and is
accompanied by a strong first-stage expansion of the doula workforce. It is best
described as credible early evidence: directionally clear and internally
consistent, but bounded by the recency of the mandates and the small number of
treated states. As the bulk of treated person-time enters the data over the next
few years, this design will provide a sharper test of whether helping hands
indeed make for healthier infants---and whether they do so most for the mothers
who need them most.

\newpage
\bibliography{references}

\begin{thebibliography}{20}
\providecommand{\natexlab}[1]{#1}
\providecommand{\url}[1]{\texttt{#1}}
\expandafter\ifx\csname urlstyle\endcsname\relax
  \providecommand{\doi}[1]{doi: #1}\else
  \providecommand{\doi}{doi: \begingroup \urlstyle{rm}\Url}\fi

\bibitem[Aizer and Currie(2014)]{aizer2014intergenerational}
Anna Aizer and Janet Currie.
\newblock The intergenerational transmission of inequality: Maternal
  disadvantage and health at birth.
\newblock \emph{Science}, 344\penalty0 (6186):\penalty0 856--861, 2014.
\newblock \doi{10.1126/science.1251872}.

\bibitem[Anderson et~al.(2020)Anderson, Brown, Charles, and
  Rees]{anderson2020occupational}
D~Mark Anderson, Ryan Brown, Kerwin~Kofi Charles, and Daniel~I Rees.
\newblock Occupational licensing and maternal health: Evidence from early
  midwifery laws.
\newblock \emph{Journal of Political Economy}, 128\penalty0 (11):\penalty0
  4337--4383, 2020.
\newblock \doi{10.1086/710555}.

\bibitem[Bohren et~al.(2017)Bohren, Hofmeyr, Sakala, Fukuzawa, and
  Cuthbert]{bohren2017continuous}
Meghan~A Bohren, G~Justus Hofmeyr, Carol Sakala, Rieko~K Fukuzawa, and Anna
  Cuthbert.
\newblock Continuous support for women during childbirth.
\newblock \emph{Cochrane Database of Systematic Reviews}, \penalty0
  (7):\penalty0 CD003766, 2017.
\newblock \doi{10.1002/14651858.CD003766.pub6}.

\bibitem[Callaway and Sant'Anna(2021)]{callaway2021difference}
Brantly Callaway and Pedro H~C Sant'Anna.
\newblock Difference-in-differences with multiple time periods.
\newblock \emph{Journal of Econometrics}, 225\penalty0 (2):\penalty0 200--230,
  2021.
\newblock \doi{10.1016/j.jeconom.2020.12.001}.

\bibitem[Cameron et~al.(2008)Cameron, Gelbach, and
  Miller]{cameron2008bootstrap}
A~Colin Cameron, Jonah~B Gelbach, and Douglas~L Miller.
\newblock Bootstrap-based improvements for inference with clustered errors.
\newblock \emph{Review of Economics and Statistics}, 90\penalty0 (3):\penalty0
  414--427, 2008.
\newblock \doi{10.1162/rest.90.3.414}.

\bibitem[Chernozhukov et~al.(2018)Chernozhukov, Chetverikov, Demirer, Duflo,
  Hansen, Newey, and Robins]{chernozhukov2018double}
Victor Chernozhukov, Denis Chetverikov, Mert Demirer, Esther Duflo, Christian
  Hansen, Whitney Newey, and James Robins.
\newblock Double/debiased machine learning for treatment and structural
  parameters.
\newblock \emph{The Econometrics Journal}, 21\penalty0 (1):\penalty0 C1--C68,
  2018.
\newblock \doi{10.1111/ectj.12097}.

\bibitem[Cygan-Rehm and Karbownik(2022)]{cyganrehm2022effects}
Kamila Cygan-Rehm and Krzysztof Karbownik.
\newblock The effects of incentivizing early prenatal care on infant health.
\newblock \emph{Journal of Health Economics}, 83:\penalty0 102612, 2022.
\newblock \doi{10.1016/j.jhealeco.2022.102612}.

\bibitem[de~Chaisemartin and
  D'Haultf{\oe}uille(2020)]{dechaisemartin2020twoway}
Cl{\'e}ment de~Chaisemartin and Xavier D'Haultf{\oe}uille.
\newblock Two-way fixed effects estimators with heterogeneous treatment
  effects.
\newblock \emph{American Economic Review}, 110\penalty0 (9):\penalty0
  2964--2996, 2020.
\newblock \doi{10.1257/aer.20181169}.

\bibitem[Falconi et~al.(2024)Falconi, Ramirez, Cobb, Levin, Nguyen, and
  Inglis]{falconi2024role}
April~M Falconi, Leah Ramirez, Rebecca Cobb, Carrie Levin, Michelle Nguyen, and
  Tiffany Inglis.
\newblock Role of doulas in improving maternal health and health equity among
  {Medicaid} enrollees, 2014--2023.
\newblock \emph{American Journal of Public Health}, 114\penalty0 (11):\penalty0
  1275--1285, 2024.
\newblock \doi{10.2105/AJPH.2024.307805}.

\bibitem[Goodman-Bacon(2021)]{goodmanbacon2021difference}
Andrew Goodman-Bacon.
\newblock Difference-in-differences with variation in treatment timing.
\newblock \emph{Journal of Econometrics}, 225\penalty0 (2):\penalty0 254--277,
  2021.
\newblock \doi{10.1016/j.jeconom.2021.03.014}.

\bibitem[Guldi and Hamersma(2023)]{guldi2023effects}
Melanie Guldi and Sarah Hamersma.
\newblock The effects of pregnancy-related {Medicaid} expansions on maternal,
  infant, and child health.
\newblock \emph{Journal of Health Economics}, 87:\penalty0 102695, 2023.
\newblock \doi{10.1016/j.jhealeco.2022.102695}.

\bibitem[Kozhimannil et~al.(2013)Kozhimannil, Hardeman, Attanasio,
  Blauer-Peterson, and O'Brien]{kozhimannil2013doula}
Katy~B Kozhimannil, Rachel~R Hardeman, Laura~B Attanasio, Cori Blauer-Peterson,
  and Michelle O'Brien.
\newblock Doula care, birth outcomes, and costs among {Medicaid} beneficiaries.
\newblock \emph{American Journal of Public Health}, 103\penalty0 (4):\penalty0
  e113--e121, 2013.
\newblock \doi{10.2105/AJPH.2012.301201}.

\bibitem[Kozhimannil et~al.(2016)Kozhimannil, Hardeman, Alarid-Escudero,
  Vogelsang, Blauer-Peterson, and Howell]{kozhimannil2016modeling}
Katy~B Kozhimannil, Rachel~R Hardeman, Fernando Alarid-Escudero, Carrie~A
  Vogelsang, Cori Blauer-Peterson, and Elizabeth~A Howell.
\newblock Modeling the cost-effectiveness of doula care associated with
  reductions in preterm birth and cesarean delivery.
\newblock \emph{Birth}, 43\penalty0 (1):\penalty0 20--27, 2016.
\newblock \doi{10.1111/birt.12218}.

\bibitem[MacKinnon and Webb(2017)]{mackinnon2017wild}
James~G MacKinnon and Matthew~D Webb.
\newblock Wild bootstrap inference for wildly different cluster sizes.
\newblock \emph{Journal of Applied Econometrics}, 32\penalty0 (2):\penalty0
  233--254, 2017.
\newblock \doi{10.1002/jae.2508}.

\bibitem[Mottl-Santiago et~al.(2023)Mottl-Santiago, Dukhovny, Cabral,
  Rodrigues, Spencer, Valle, and Feinberg]{mottlsantiago2023effectiveness}
Julie Mottl-Santiago, Dmitry Dukhovny, Howard Cabral, Dona Rodrigues, Linda
  Spencer, Eduardo~A Valle, and Emily Feinberg.
\newblock Effectiveness of an enhanced community doula intervention in a safety
  net setting: A randomized controlled trial.
\newblock \emph{Health Equity}, 7\penalty0 (1):\penalty0 466--476, 2023.
\newblock \doi{10.1089/heq.2022.0200}.

\bibitem[Noghanibehambari and Fletcher(2023)]{noghani2023longterm}
Hamid Noghanibehambari and Jason Fletcher.
\newblock Long-term health benefits of occupational licensing: Evidence from
  midwifery laws.
\newblock \emph{Journal of Health Economics}, 92:\penalty0 102807, 2023.
\newblock \doi{10.1016/j.jhealeco.2023.102807}.

\bibitem[Peet et~al.(2022)Peet, Schultz, Lovejoy, and Tsui]{peet2022variation}
Evan~D Peet, Dana Schultz, Susan Lovejoy, and Fuchiang Tsui.
\newblock Variation in the infant health effects of the women, infants, and
  children program by predicted risk using novel machine learning methods.
\newblock \emph{Health Economics}, 31\penalty0 (12):\penalty0 2604--2615, 2022.
\newblock \doi{10.1002/hec.4617}.

\bibitem[Peet et~al.(2024)Peet, Schultz, and Tsui]{peet2024infant}
Evan~D Peet, Dana Schultz, and Fuchiang Tsui.
\newblock The infant health effects of doulas: Leveraging big data and machine
  learning to inform cost-effective targeting.
\newblock \emph{Health Economics}, 33\penalty0 (6):\penalty0 1387--1411, 2024.
\newblock \doi{10.1002/hec.4821}.

\bibitem[Rambachan and Roth(2023)]{rambachan2023more}
Ashesh Rambachan and Jonathan Roth.
\newblock A more credible approach to parallel trends.
\newblock \emph{Review of Economic Studies}, 90\penalty0 (5):\penalty0
  2555--2591, 2023.
\newblock \doi{10.1093/restud/rdad018}.

\bibitem[Sonchak(2015)]{sonchak2015medicaid}
Lyudmyla Sonchak.
\newblock Medicaid reimbursement, prenatal care and infant health.
\newblock \emph{Journal of Health Economics}, 44:\penalty0 10--24, 2015.
\newblock \doi{10.1016/j.jhealeco.2015.08.008}.

\end{thebibliography}

\newpage
\appendix
\section{Appendix: Additional Figures}\label{app:figures}

\begin{figure}[H]
\centering
\includegraphics[width=0.49\linewidth]{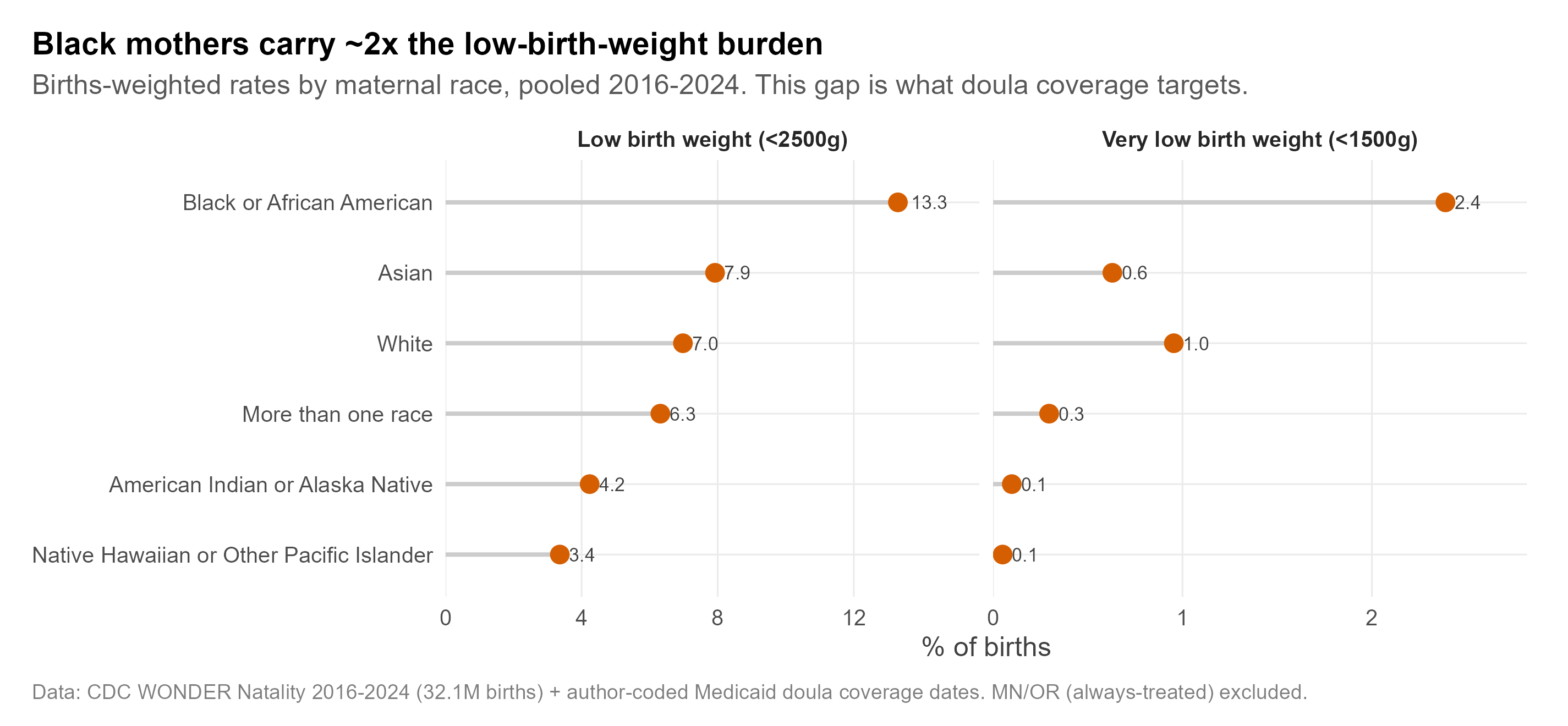}\hfill
\includegraphics[width=0.49\linewidth]{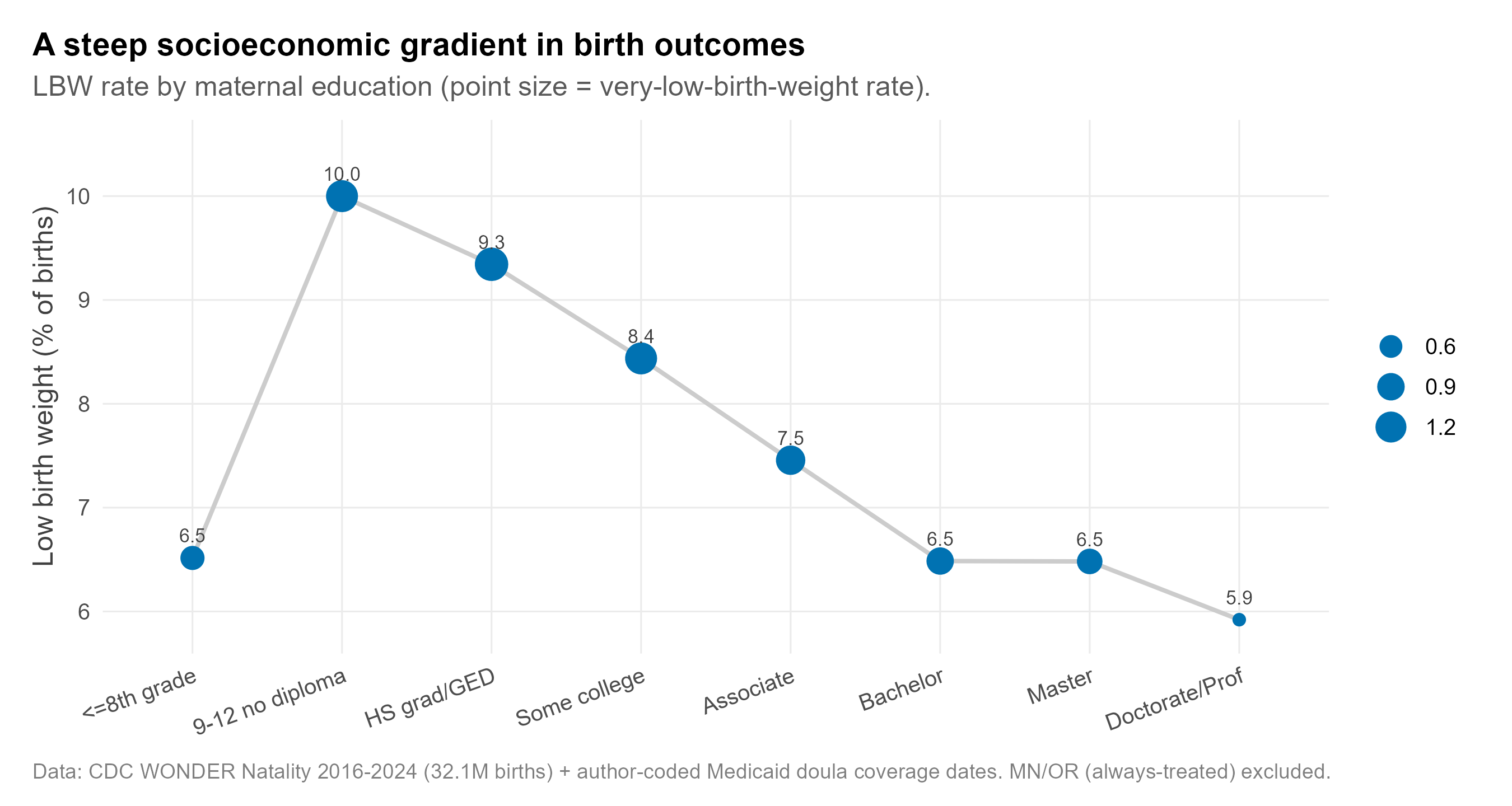}
\caption{Left: LBW and VLBW by maternal race (pooled). Right: the socioeconomic
gradient in LBW by maternal education.}
\label{fig:descr}
\end{figure}

\begin{figure}[H]
\centering
\includegraphics[width=0.49\linewidth]{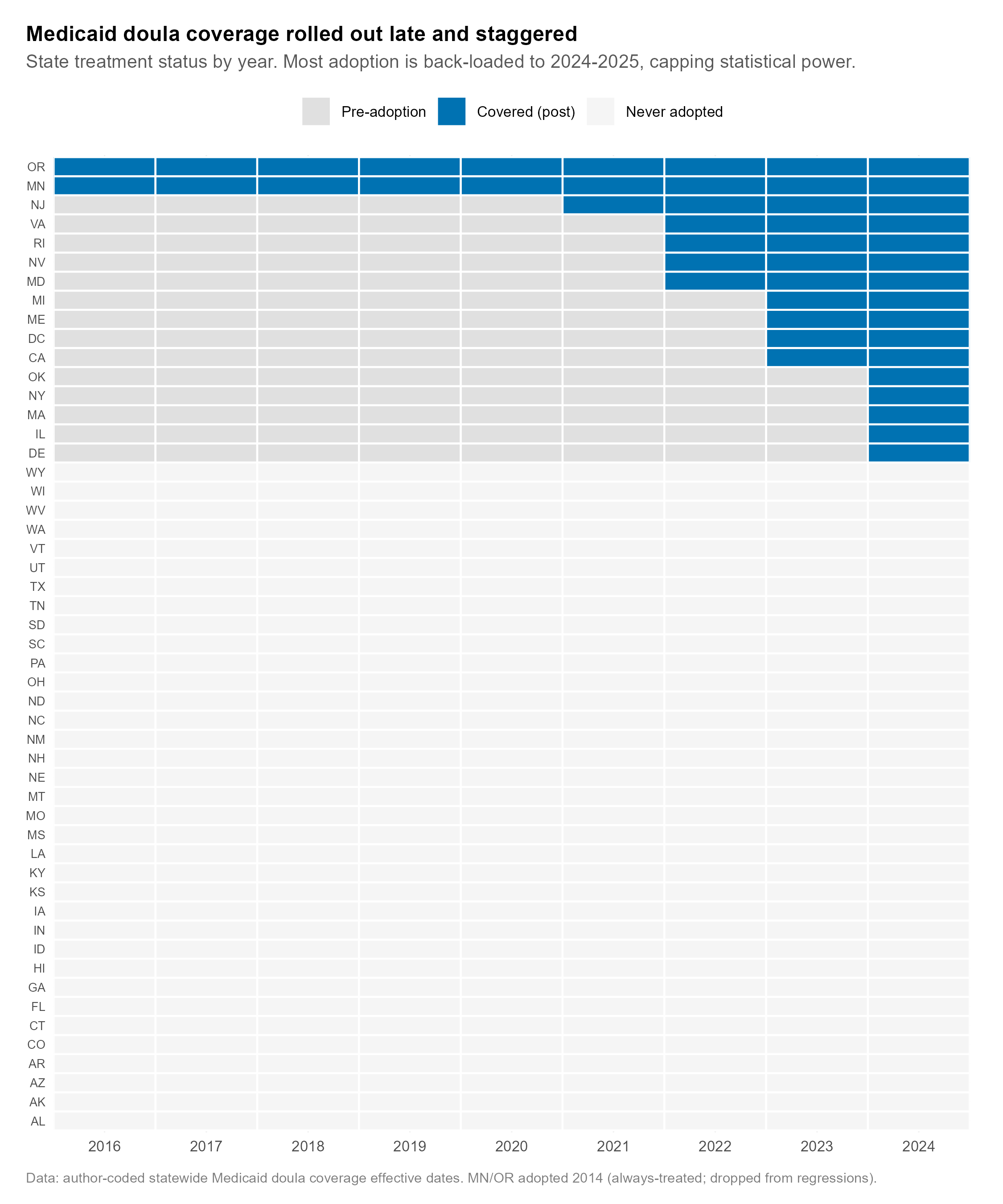}\hfill
\includegraphics[width=0.49\linewidth]{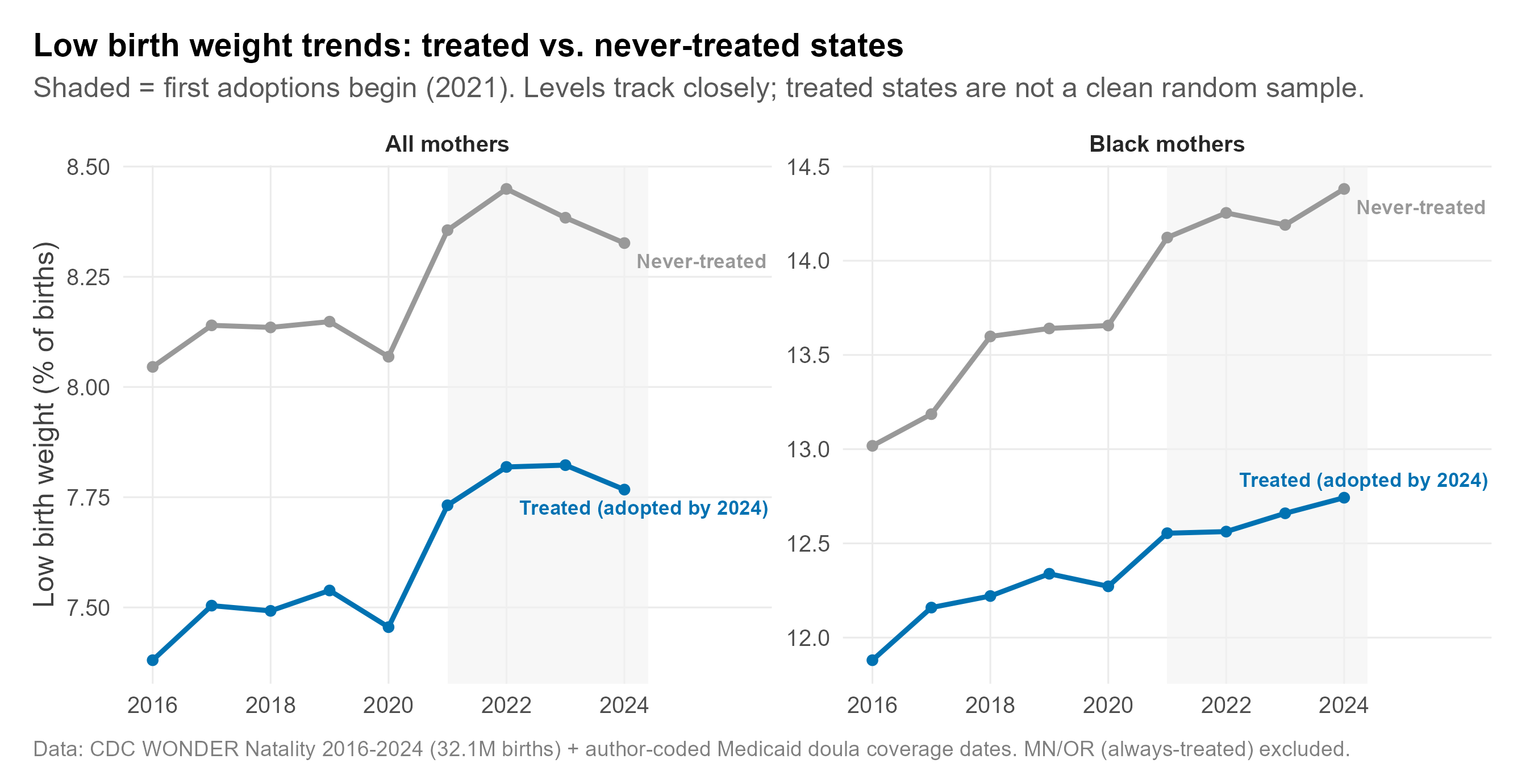}
\caption{Left: staggered adoption (state $\times$ year). Right: LBW trends,
treated vs.\ never-treated, all mothers and Black mothers.}
\label{fig:rollout}
\end{figure}

\begin{figure}[H]
\centering
\includegraphics[width=0.49\linewidth]{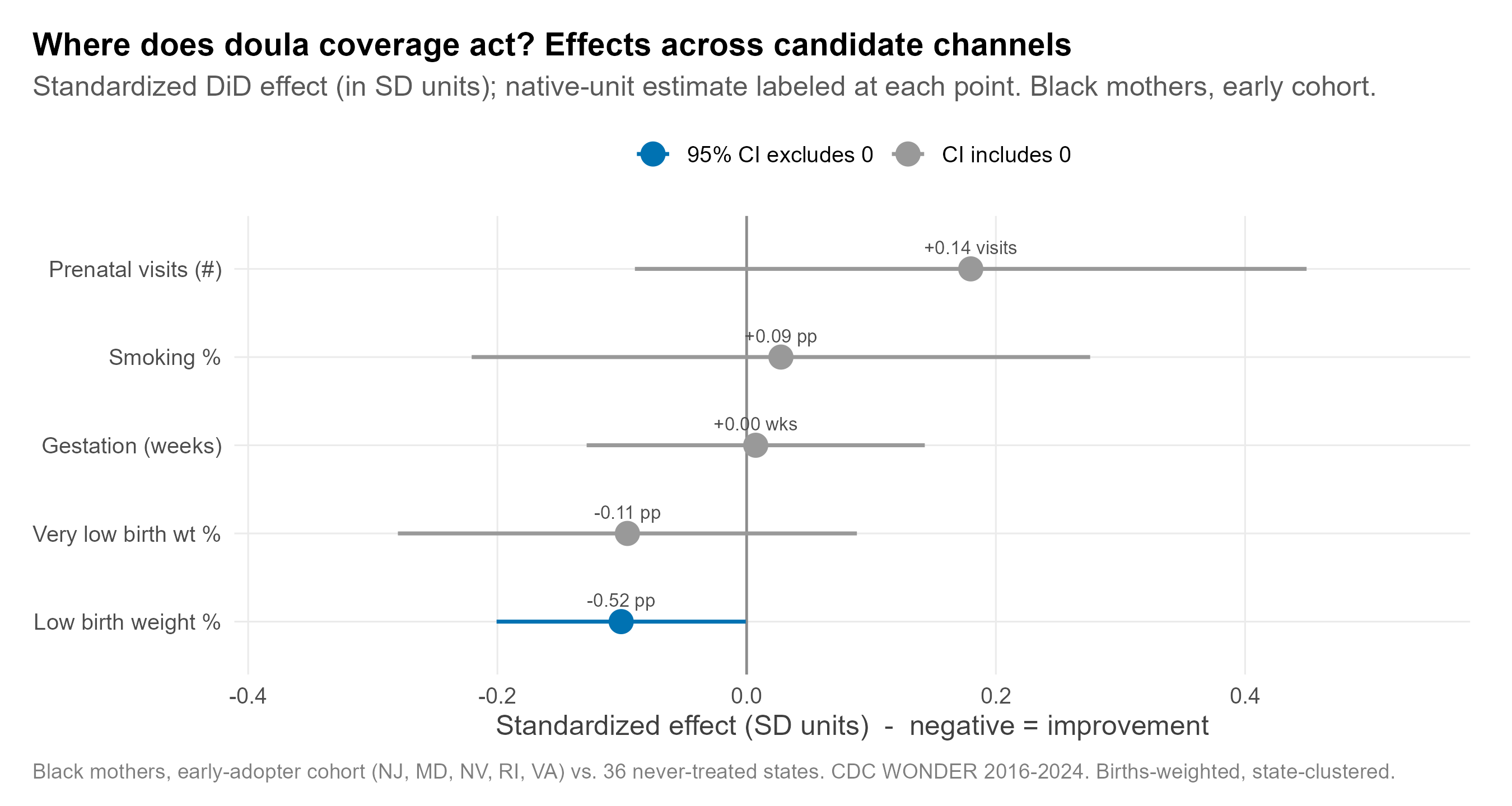}\hfill
\includegraphics[width=0.49\linewidth]{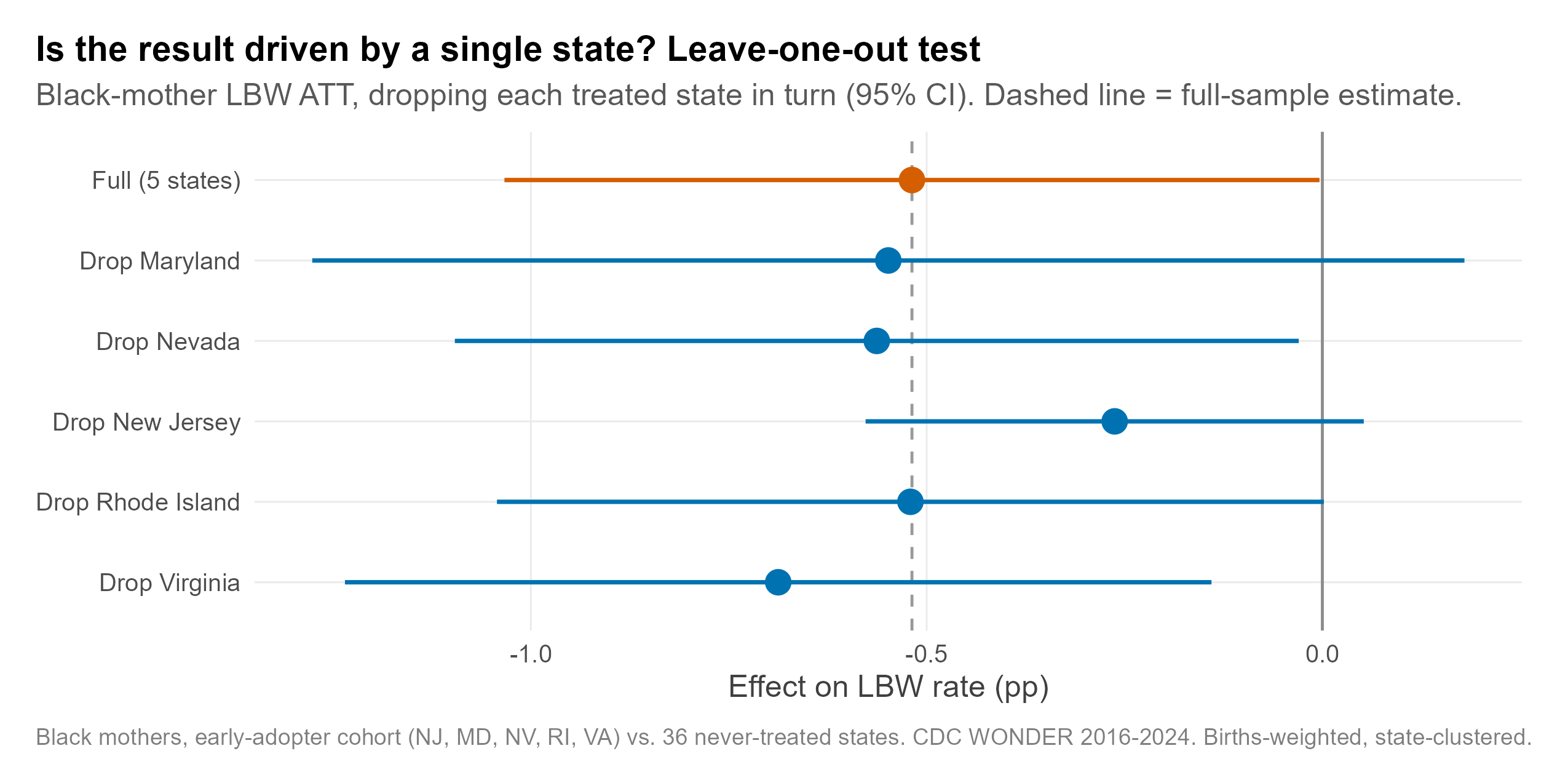}
\caption{Left: effects across candidate channels for Black mothers (only LBW
moves). Right: leave-one-state-out robustness of the Black-mother estimate.}
\label{fig:channels}
\end{figure}

\begin{figure}[H]
\centering
\includegraphics[width=0.6\linewidth]{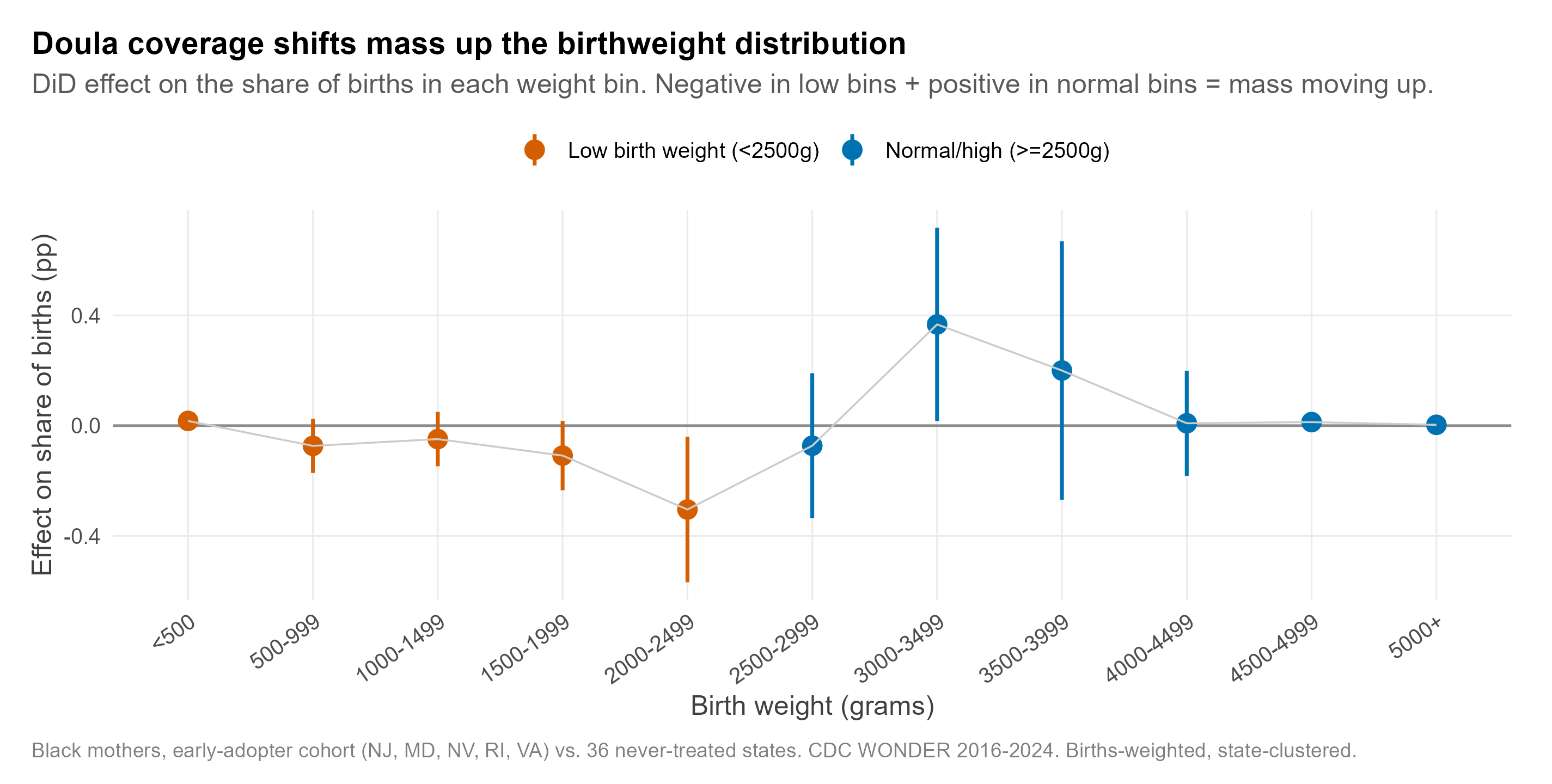}
\caption{TWFE estimates across the birth-weight distribution for Black mothers
(early cohort), shown for comparison with the Callaway--Sant'Anna version in
Figure~\ref{fig:cs}.}
\label{fig:twfedist}
\end{figure}

\begin{figure}[H]
\centering
\includegraphics[width=0.98\linewidth]{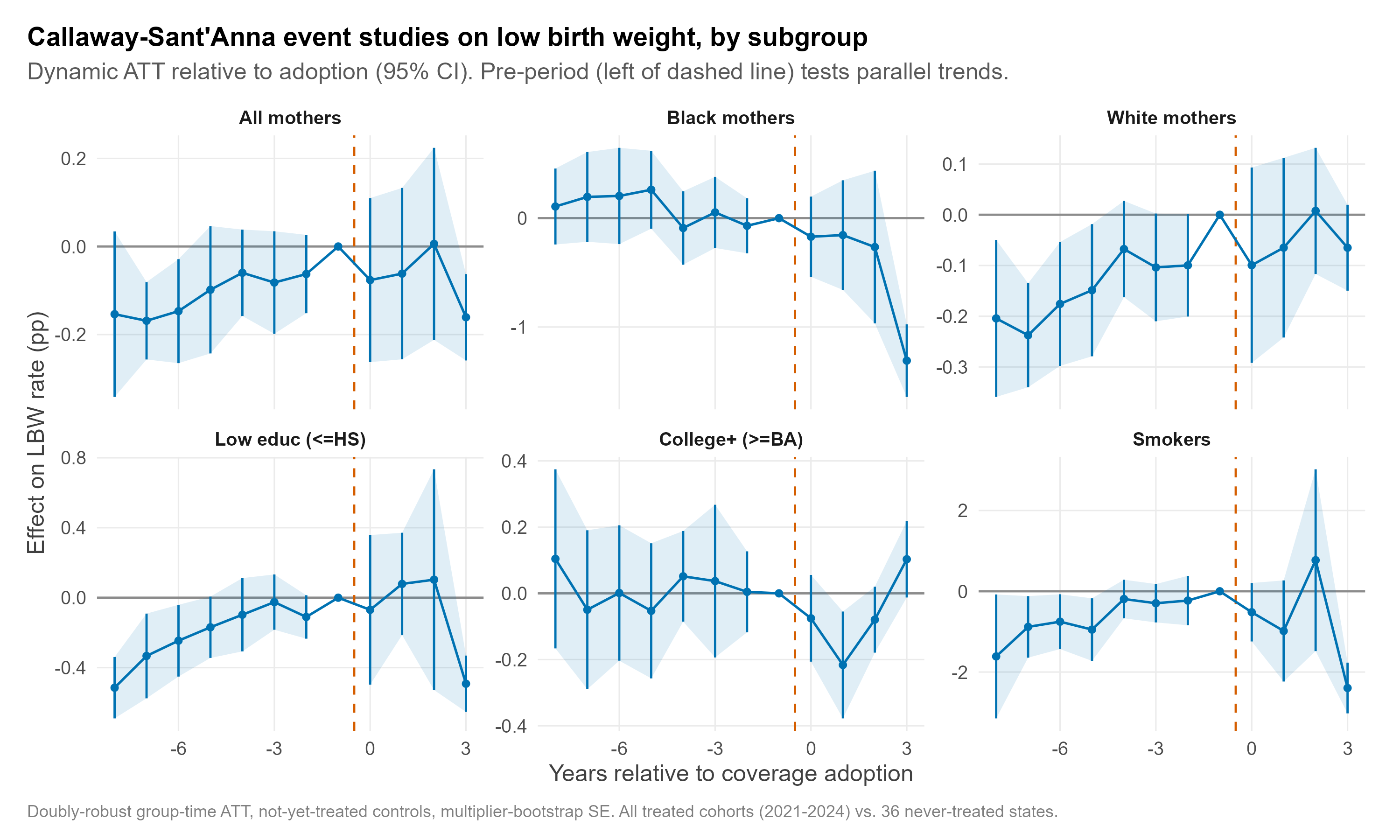}
\caption{Callaway--Sant'Anna event-study estimates on low birth weight for every
subgroup (dynamic ATT relative to adoption, 95\% CI; all treated cohorts vs.\
never-treated). The pre-period (left of the dashed line) tests parallel trends:
it is flat for Black mothers, whereas other subgroups show pre-adoption movement,
consistent with the pre-trend tests in Table~\ref{tab:hetero}.}
\label{fig:cssub}
\end{figure}

\begin{table}[H]
\centering
\caption{Callaway--Sant'Anna estimates across the birth-weight distribution (Black mothers)}
\label{tab:csbins}
\begin{threeparttable}
\begin{tabular}{lcc}
\toprule
Birth-weight bin (grams) & ATT (pp) & SE \\
\midrule
$<$500          & $0.018$        & 0.019 \\
500--999        & $0.023$        & 0.042 \\
1{,}000--1{,}499& $-0.004$       & 0.052 \\
1{,}500--1{,}999& $-0.080^{**}$  & 0.035 \\
2{,}000--2{,}499& $-0.187$       & 0.150 \\
2{,}500--2{,}999& $-0.019$       & 0.140 \\
3{,}000--3{,}499& $\phantom{-}0.330^{**}$ & 0.156 \\
3{,}500--3{,}999& $-0.038$       & 0.192 \\
4{,}000--4{,}499& $-0.043$       & 0.105 \\
4{,}500--4{,}999& $-0.002$       & 0.035 \\
5{,}000$+$      & $\phantom{-}0.002$ & 0.006 \\
\bottomrule
\end{tabular}
\begin{tablenotes}\footnotesize
\item Effect on the share of Black births in each weight bin (percentage points).
Callaway--Sant'Anna group-time ATT, doubly robust, not-yet-treated comparison
group, multiplier-bootstrap SE. $^{**}$ indicates the 95\% CI excludes zero. Mass
shifts out of the 1{,}500--1{,}999g bin and into the 3{,}000--3{,}499g range.
\end{tablenotes}
\end{threeparttable}
\end{table}

\end{document}